\documentclass[fleqn,10pt]{wlscirep}
\usepackage[utf8]{inputenc}
\usepackage[T1]{fontenc}
\usepackage{multirow}
\usepackage{makecell}
\usepackage{xcolor}
\usepackage{subcaption}
\usepackage{colortbl}

\title{Alignment between Brains and AI: Evidence for Convergent Evolution across Modalities, Scales \\and Training Trajectories}

\author[1,3,$\dagger$]{Guobin Shen}
\author[1,4,$\dagger$]{Dongcheng Zhao}
\author[1,3]{Yiting Dong}
\author[1,4]{Qian Zhang}
\author[1,2,4,*]{Yi Zeng} 

\affil[1]{Laboratory of Brain-inspired Cognitive AI, Institute of Automation, Chinese Academy of Sciences, Beijing, 100190, China}
\affil[2]{State Key Laboratory of Brain Cognition and Brain-inspired Intelligence Technology, Chinese Academy of Sciences, Shanghai, 200031, China}
\affil[3]{School of Future Technology, University of Chinese Academy of Sciences, Beijing, 100049, China}
\affil[4]{Center for Long-term AI, Beijing, 101407, China}

\affil[*]{corresponding author: Yi Zeng (yi.zeng@ia.ac.cn)}
\affil[$\dagger$]{these authors contributed equally to this work}

\begin{abstract} 
    Artificial and biological systems may evolve similar computational solutions despite fundamental differences in architecture and learning mechanisms—a form of convergent evolution. We demonstrate this phenomenon through large-scale analysis of alignment between human brain activity and internal representations of over 600 AI models spanning language and vision domains, from 1.33M to 72B parameters. Analyzing 60 million alignment measurements reveals that higher-performing models spontaneously develop stronger brain alignment without explicit neural constraints, with language models showing markedly stronger correlation ($r=0.89$, $p<7.5 \times 10^{-13}$) than vision models ($r=0.53$, $p<2.0 \times 10^{-44}$). Crucially, longitudinal analysis demonstrates that brain alignment consistently precedes performance improvements during training, suggesting that developing brain-like representations may be a necessary stepping stone toward higher capabilities. We find systematic patterns: language models exhibit strongest alignment with limbic and integrative regions, while vision models show progressive alignment with visual cortices; deeper processing layers converge across modalities; and as representational scale increases, alignment systematically shifts from primary sensory to higher-order associative regions. These findings provide compelling evidence that optimization for task performance naturally drives AI systems toward brain-like computational strategies, offering both fundamental insights into principles of intelligent information processing and practical guidance for developing more capable AI systems.
\end{abstract}

\begin{document}

\flushbottom
\maketitle
%
%
\thispagestyle{empty}


\section*{Introduction}

The rapid progress of Artificial Intelligence (AI) has raised a fundamental question: as these systems increasingly match or surpass human-level performance in language, vision, and reasoning domains~\cite{lecun2015deep, silver2016mastering, brown2020language}, do their internal representations also converge toward brain-like computational strategies? This question becomes more pressing as AI-generated outputs become indistinguishable from human-created content in interactive settings~\cite{jones2025large}. Understanding these internal representational mechanisms is increasingly urgent both for advancing our knowledge of intelligence and for guiding the development of safer, more interpretable AI systems.

Recent advances in Large Language Models (LLMs)~\cite{brown2020language, grattafiori2024llama, yang2024qwen2} and vision models~\cite{dosovitskiy2020image, bommasani2021opportunities} provide an unprecedented opportunity to empirically test this convergent evolution hypothesis. Prior studies have demonstrated that deep neural networks map onto primate visual cortex hierarchies~\cite{yamins2014performance} or transformer-based LLMs align with human language areas~\cite{schrimpf2021neural}. However, existing work has been constrained by three key limitations: analyses restricted to single modalities, examination of narrow cortical regions rather than distributed networks, and reliance on static model checkpoints that fail to capture representational evolution during training~\cite{storrs2021diverse, zhuang2017toward}. These constraints have prevented a comprehensive understanding of whether and how AI systems converge toward brain-like representations across diverse computational domains.

In both biological and artificial systems, internal representations refer to distributed patterns of activity that encode information relevant for behavior and cognition. In the brain, these representations are shaped by experience and task demands, forming hierarchical organizations from sensory to conceptual levels~\cite{yamins2016using}. Deep learning models exhibit analogous properties, with internal states evolving across layers to capture increasingly abstract features. Examining these internal dynamics beyond merely analyzing outputs uncovers fundamental principles of intelligent information processing. This approach illuminates the mechanisms underlying the emergence of brain-like computational strategies, providing insights into the fundamental nature of intelligence.

Convergent evolution provides a powerful framework for understanding brain-AI alignment. Just as vertebrates and cephalopods independently evolved camera-like eyes under similar visual processing demands~\cite{nilsson2013eye, ogura2004comparative}, artificial and biological systems may converge on similar computational strategies when facing shared information-processing challenges~\cite{zador2019critique, lillicrap2020backpropagation, hassabis2017neuroscience}. This convergence has profound implications: high-performing AI models offer testbeds for understanding biological computation~\cite{richards2019deep, hassabis2017neuroscience}, while brain organization may guide development of more robust and efficient AI systems~\cite{zador2019critique}. Brain-AI alignment thus serves as both a diagnostic tool and a blueprint for future intelligence architectures.

To systematically test this convergent evolution hypothesis and address the limitations of prior work, we present a comprehensive analysis of brain-AI alignment unprecedented in scale. Our study examines internal representations from over 600 models across diverse architectures, scales, training trajectories, and modalities. Through analyzing 60 million alignment measurements, we directly compare layer-wise activations to human neural recordings, addressing three fundamental questions: (1) Does brain alignment precede performance improvements during training, suggesting it may be a necessary stepping stone? (2) Do alignment patterns differ systematically between vision and language modalities, or converge toward universal principles? (3) How do model hierarchies progressively map onto cortical processing levels from sensory to associative regions? Our findings reveal that artificial and biological intelligence, despite their distinct evolutionary paths, indeed converge toward similar computational solutions. This establishes fundamental principles that could transform both our understanding of intelligence and the development of future AI systems.

\section*{Results}

We developed a systematic large-scale framework to comprehensively assess brain-AI representational alignment across sensory modalities, model scales, and training dynamics (Figure~\ref{fig:performance_alignment}). Our analysis leveraged fMRI recordings from the Natural Scenes Dataset (NSD)~\cite{allen2022massive}, which captures neural activity from multiple subjects viewing thousands of naturalistic images. These images, sourced from the COCO dataset~\cite{lin2014microsoft} and paired with human-generated captions, enabled multimodal analyses across both vision and language domains.

Our AI model collection comprised 630 neural networks spanning diverse architectures and scales: 36 large language models (0.5-72B parameters) including Qwen~\cite{yang2024qwen2}, Llama~\cite{grattafiori2024llama}, and Gemma~\cite{team2024gemma} families, and 594 vision models (1.33-1014M parameters) from CNNs to transformers~\cite{krizhevsky2012imagenet, dosovitskiy2020image}. To quantify alignment, we employed Centered Kernel Alignment (CKA)~\cite{kornblith2019similarity} across multiple spatial scales, mapping model representations to brain regions defined by the HCP\_MMP1 parcellation~\cite{glasser2016multi} and Yeo-7 functional networks~\cite{yeo2011organization}. For longitudinal analyses, we tracked representational evolution in the Pythia language model family~\cite{biderman2023pythia} and MixNet vision model family~\cite{tan2019mixconv} throughout training.

\subsection*{Correlation Patterns Between Brain Alignment and Model Effectiveness}

\begin{figure}[h!]
    \centering
    \includegraphics[width=1.0\textwidth]{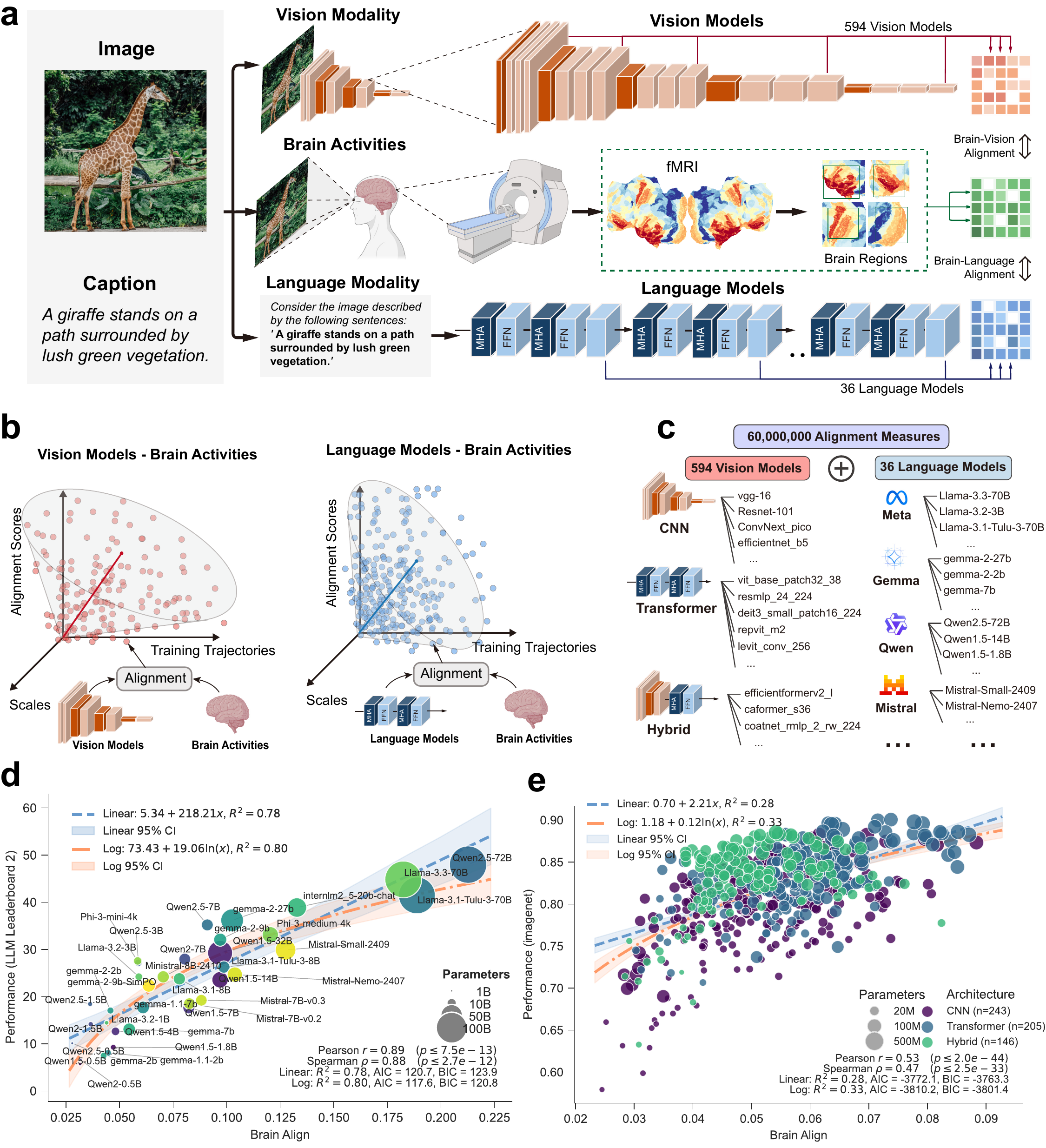}

    \vspace{-.36cm}
    \caption{\textbf{Relationship between model performance and brain alignment.} \textbf{a}, Methodological overview showing the multi-modal analysis framework for vision models, brain recordings, and language models. \textbf{b}, Multi-dimensional analytical approach spanning scales, training trajectories, and modalities for Brain-AI alignment. \textbf{c}, Comprehensive analysis of 594 vision models and 36 language models, enabling 60 million distinct alignment measurements. \textbf{d}, Language models show strong correlation between LLM Leaderboard 2~\cite{open-llm-leaderboard-v2} performance and brain alignment scores, with logarithmic fit outperforming linear fit. Parameter counts indicated by marker size. \textbf{e}, Vision models demonstrate moderate but significant correlation between ImageNet top-1 accuracy and brain alignment, with differentiation between architectural types.}
    \label{fig:performance_alignment}
    \vspace{-1cm}
\end{figure}
    
Our first key finding is a robust, positive correlation between model performance and brain alignment across both language and vision domains (Figure~\ref{fig:performance_alignment}d-e). For performance evaluation, we utilized two widely recognized benchmarks: the LLM Leaderboard 2~\cite{open-llm-leaderboard-v2} for language models, which aggregates performance across multiple reasoning, knowledge, and language understanding tasks into a composite score, and ImageNet~\cite{deng2009imagenet} Top-1 accuracy for vision models, which measures classification performance on a diverse dataset of 1,000 object categories. For language models, brain alignment showed a strong correlation with performance (Pearson $r = 0.89$, $p < 7.5 \times 10^{-13}$; Spearman $\rho = 0.88$, $p < 2.7 \times 10^{-12}$), with the relationship better characterized by a logarithmic fit ($R^2 = 0.80$, $AIC = 117.6$, $BIC = 120.8$) than a linear one ($R^2 = 0.78$, $AIC = 120.7$, $BIC = 123.9$). This pattern held across different model families, with higher-performing models generally showing stronger brain alignment.

The logarithmic relationship suggests a principled trajectory of diminishing returns, where initial improvements in performance correspond to substantial gains in brain alignment, followed by more modest alignment improvements as performance continues to increase. Notably, we observed that language models with similar performance but different architectural designs and fine-tuning approaches showed distinct brain alignment scores. For instance, Llama-3.1-Tulu-3-8B~\cite{lambert2024t} exhibits higher brain alignment and performance than the base Llama-3.1-8B~\cite{grattafiori2024llama} model, suggesting that certain fine-tuning approaches may enhance brain-like representations. Similarly, gemma-2-9b-SimPO~\cite{meng2024simpo} shows lower brain alignment and performance compared to the standard gemma-2-9b~\cite{team2024gemma}, indicating that different training methodologies impact alignment with biological representations even within the same model family.

Similarly, for vision models evaluated on ImageNet, we observed a significant correlation between accuracy and brain alignment (Pearson $r = 0.53$, $p < 2.0 \times 10^{-44}$; Spearman $\rho = 0.47$, $p < 2.5 \times 10^{-33}$). The correlation was consistent across model sizes, from small models (under 10M parameters) to large models (over 1000M parameters), indicating a scale-invariant relationship between performance optimization and brain alignment. An important observation in the vision domain is that transformer-based vision models exhibited slightly higher brain alignment than CNNs at equivalent performance levels, suggesting that self-attention mechanisms may naturally encourage more brain-like representations. 

Both vision and language models exhibit a logarithmic relationship between performance and brain alignment, revealing an interesting pattern of diminishing returns. As model performance approaches benchmark saturation, brain alignment scores continue to increase. This suggests models may develop increasingly brain-like representations even after task performance reaches ceiling effects on standard benchmarks.

\begin{table}[ht]
    \centering
    \renewcommand{\arraystretch}{1.3}  
    \setlength{\tabcolsep}{5pt}        
    \begin{tabular}{l|l|cccc|cccc} 
    \hline
    \multicolumn{2}{c|}{\textbf{Model Information}} & 
    \multicolumn{4}{c|}{\textbf{Pearson Correlation}} & 
    \multicolumn{4}{c}{\textbf{Spearman Correlation}} \\ 
    \hline
    \multicolumn{1}{c|}{\textbf{Model}} & 
    \multicolumn{1}{c|}{\textbf{Benchmark}} & 
    \multicolumn{1}{c|}{\textbf{$r$}} & 
    \multicolumn{1}{c|}{\textbf{95\% CI}} & 
    \multicolumn{1}{c|}{\textbf{$p$-value}} & 
    \multicolumn{1}{c|}{\textbf{FDR p}} & 
    \multicolumn{1}{c|}{\textbf{$\rho$}} & 
    \multicolumn{1}{c|}{\textbf{95\% CI}} & 
    \multicolumn{1}{c|}{\textbf{$p$-value}} & 
    \multicolumn{1}{c}{\textbf{FDR $p$}} \\ 
    \hline
    \multirow{9}{*}{\makecell[l]{Language\\Models}} 
    & IFEval~\cite{zhou2023instruction} & 0.77 & [0.59, 0.88] & <3.5e-08 & <4.1e-08 & 0.79 & [0.58, 1.00] & <1.3e-08 & <1.6e-08 \\ \cline{2-10}
    & BBH~\cite{suzgun2023challenging} & 0.84 & [0.70, 0.91] & <2.3e-10 & <3.0e-10 & 0.86 & [0.65, 1.07] & <1.9e-11 & <2.7e-11 \\ \cline{2-10}
    & MATH Lvl 5~\cite{hendrycks2measuring} & 0.76 & [0.58, 0.87] & <7.6e-08 & <8.7e-08 & 0.67 & [0.46, 0.88] & <7.1e-06 & <7.6e-06 \\ \cline{2-10}
    & GPQA~\cite{rein2024gpqa} & 0.76 & [0.57, 0.87] & <7.7e-08 & <8.7e-08 & 0.78 & [0.57, 0.99] & <2.3e-08 & <2.8e-08 \\ \cline{2-10}
    & MUSR~\cite{sprague2024musr} & 0.66 & [0.42, 0.81] & <1.3e-05 & <1.3e-05 & 0.59 & [0.38, 0.80] & <1.6e-04 & <1.6e-04 \\ \cline{2-10}
    & \makecell[l]{MMLU-PRO~\cite{wang2024mmlu}} & 0.83 & [0.69, 0.91] & <3.8e-10 & <5.0e-10 & 0.85 & [0.64, 1.06] & <7.3e-11 & <1.0e-10 \\ \cline{2-10}
    & Leaderboard 2~\cite{open-llm-leaderboard-v2} & 0.89 & [0.79, 0.94] & <7.5e-13 & <1.1e-12 & 0.88 & [0.67, 1.09] & <2.7e-12 & <4.0e-12 \\ \cline{2-10}
    & Chatbot Arena~\cite{chiang2024chatbot} & 0.73 & [0.48, 0.87] & <2.3e-05 & <2.4e-05 & 0.80 & [0.55, 1.05] & <9.9e-07 & <1.1e-06 \\ \hline
    \multirow{10}{*}{\makecell[l]{Vision\\Models}} 
    & ImageNet~\cite{deng2009imagenet} & 0.53 & [0.47, 0.59] & <2.0e-44 & <1.5e-43 & 0.47 & [0.41, 0.52] & <2.5e-33 & <6.5e-33 \\ \cline{2-10}
    & \makecell[l]{ImageNet-A~\cite{hendrycks2021natural}} & 0.46 & [0.39, 0.52] & <7.5e-32 & <1.7e-31 & 0.45 & [0.40, 0.50] & <2.1e-31 & <4.4e-31 \\ \cline{2-10}
    & \makecell[l]{ImageNet-A-Clean~\cite{hendrycks2021natural}} & 0.51 & [0.45, 0.57] & <1.7e-41 & <7.0e-41 & 0.46 & [0.41, 0.51] & <2.3e-32 & <5.6e-32 \\ \cline{2-10}
    & \makecell[l]{ImageNet-R~\cite{hendrycks2021many}} & 0.52 & [0.46, 0.58] & <8.3e-43 & <4.6e-42 & 0.47 & [0.42, 0.52] & <1.1e-33 & <3.1e-33 \\ \cline{2-10}
    & \makecell[l]{ImageNet-R-Clean~\cite{hendrycks2021natural}} & 0.51 & [0.45, 0.57] & <2.8e-40 & <1.0e-39 & 0.45 & [0.40, 0.50] & <5.5e-31 & <1.1e-30 \\ \cline{2-10}
    & \makecell[l]{ImageNet-Real~\cite{beyer2020we}} & 0.50 & [0.44, 0.56] & <1.2e-38 & <4.0e-38 & 0.44 & [0.39, 0.49] & <2.9e-29 & <5.0e-29 \\ \cline{2-10}
    & \makecell[l]{ImageNetv2-\\matched-freq.~\cite{recht2019imagenet}} & 0.52 & [0.46, 0.57] & <5.6e-42 & <2.4e-41 & 0.45 & [0.40, 0.50] & <1.3e-30 & <2.4e-30 \\ \cline{2-10}
    & sketch~\cite{wang2019learning} & 0.53 & [0.47, 0.58] & <1.3e-43 & <8.2e-43 & 0.47 & [0.42, 0.52] & <4.0e-34 & <1.2e-33 \\ \hline
    \end{tabular}
    \caption{\textbf{Correlation between model performance and brain alignment across benchmarks.} Results show correlation metrics across different performance benchmarks for language and vision models. FDR-corrected p-values account for multiple comparisons.}
    \label{tab:correlation_stats}
\end{table}

Table~\ref{tab:correlation_stats} presents detailed correlation statistics between brain alignment and performance across multiple benchmarks. For language models, the composite LLM Leaderboard 2 metric showed the strongest correlation ($r=0.89$), with component benchmarks exhibiting similarly robust relationships: BBH ($r=0.84$), MMLU-PRO ($r=0.83$), IFEval ($r=0.77$), MATH Level 5 ($r=0.76$), and GPQA ($r=0.76$). Notably, Chatbot Arena~\cite{chiang2024chatbot}, which relies on human evaluation of conversational quality, showed a slightly lower correlation ($r=0.73$). This difference likely reflects that our measurement paradigm aligns more closely with objective capability assessment than with the subjective criteria underlying human conversational preferences. Vision models demonstrated moderate but consistent correlations across evaluation contexts, from standard ImageNet ($r=0.53$) to more challenging variants like ImageNet-R ($r=0.52$) and ImageNet-A ($r=0.46$). This pattern suggests that higher-level cognitive tasks, particularly those requiring complex reasoning abilities, may more strongly drive representational convergence between artificial and biological systems. Additional visualizations showing performance-brain alignment correlations for all metrics listed in Table~\ref{tab:correlation_stats} are provided in the supplementary materials.

The consistent brain-performance correlation across vastly different model architectures, training objectives, and modalities provides compelling evidence for a form of convergent evolution, whereby optimization for task performance naturally leads AI systems toward more brain-like representations without explicit neurobiological constraints. The substantially stronger correlations observed in language models compared to vision models likely reflects that language processing involves more abstract and complex cognitive operations, such as semantic integration, contextual reasoning, and compositional understanding, which may naturally drive both biological and artificial systems toward similar computational solutions. This modality gap in correlation strength is maintained across different benchmarks and evaluation contexts, suggesting that as tasks become more cognitively demanding, artificial systems increasingly converge with biological processing strategies.

Importantly, these alignment patterns were highly consistent across different subjects, as demonstrated by the strong inter-subject correlations in both vision and language modalities. Our comprehensive consistency analysis verified this robustness (Tables~\ref{tab:lang_corr},~\ref{tab:vision_corr}), with inter-subject correlations ranging from 0.997 to 1.000 for language models and 0.970 to 0.999 for vision models. These remarkably high cross-subject correlations indicate that the observed representational correspondences reflect universal computational principles rather than individual variations, suggesting that both biological and artificial intelligence systems converge on similar solutions when facing the same information processing challenges.

\subsection*{Architectural Correspondences Between Brains and AI}

\begin{figure}[hb]
    \centering
    \includegraphics[width=1.0\textwidth]{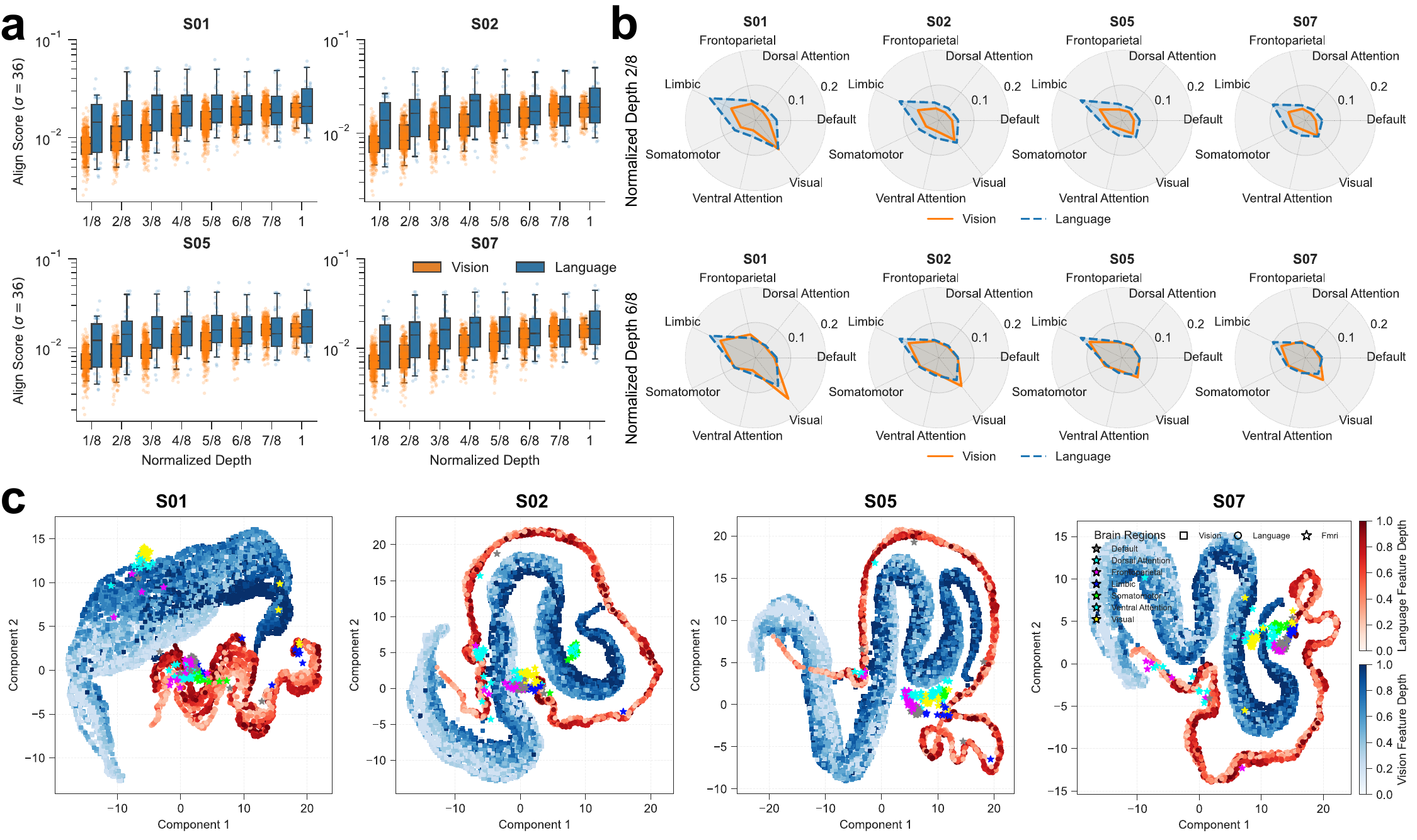}
    \vspace{-.36cm} 
    \caption{\textbf{Layer-wise brain alignment patterns across vision and language models.} \textbf{a}, Brain alignment by normalized layer depth for vision and language models across subjects. For consistent comparison across models with different architectures, each model's layers were normalized into 8 depth segments (1/8 through 8/8). \textbf{b}, Comparison of brain network alignment at normalized depths 2/8 and 6/8 for vision versus language models across Yeo-7 brain networks~\cite{yeo2011organization}. \textbf{c}, UMAP~\cite{mcinnes2018umap} visualization of 2D embedding space based on brain region alignment patterns, showing relationships between model layers and brain regions. Each point represents either a model layer or brain region, with color intensity indicating the depth for model layers, and different colors representing distinct brain regions.}
    \label{fig:layer_alignment}     
\end{figure}


We next examined how brain alignment varies across layers of deep neural networks. To enable consistent comparisons across architectures of varying depth, we divided each model into 8 normalized layer segments (from 1/8 to 8/8) and averaged alignment scores within each. As shown in Figure~\ref{fig:layer_alignment}a, vision models exhibited a gradual increase in alignment with depth, peaking at the final segments (6/8 to 8/8). In contrast, language models showed overall higher brain alignment, with a distinct middle-layer peak (3/8 to 5/8) that remained relatively stable through deeper layers.

When examining alignment with specific brain networks defined by the Yeo-7~\cite{yeo2011organization} parcellation (Figure~\ref{fig:layer_alignment}b), we observed clear modality-specific patterns. At normalized depth 2/8 (shallow layers), vision models showed modest alignment across all networks with a slight preference for visual regions, while language models demonstrated stronger overall alignment with pronounced correspondence to limbic and default mode networks. At normalized depth 6/8 (deeper layers), vision models developed substantially stronger alignment with visual network regions while maintaining lower alignment to other networks. Language models at this deeper level showed a more balanced pattern across networks, maintaining strong alignment to limbic and default regions. This distinct pattern was consistent across all subjects, suggesting a fundamental organizational principle rather than individual variation.

Interestingly, language models showed substantial alignment with non-language-specific brain regions associated with abstract concept representation and semantic integration. This pattern suggests that language models may capture general-purpose computational principles employed across diverse cognitive domains rather than language-specific representations alone. Vision models, conversely, showed more modality-specific alignment concentrated in visual processing regions.

Dimensionality reduction analysis further illuminated these representational relationships (Figure~\ref{fig:layer_alignment}c). Using UMAP~\cite{mcinnes2018umap} to project the high-dimensional alignment scores into a 2D space revealed two distinct manifolds: vision (blue squares) and language (red circles) model representations, with color intensity indicating layer depth progression from early to deeper processing stages. Brain regions (represented as stars) are derived from the HCP\_MMP1~\cite{glasser2016multi} parcellation and are colored according to their Yeo-7 network affiliation. These brain regions are distributed across the space based on their alignment patterns with different model layers, providing a visual map of how neural and artificial representations relate to each other.

Notably, deeper vision model layers positioned closer to the language model manifold, suggesting that as visual processing becomes more abstract, it begins to share representational properties with language processing. Brain regions followed a similar organizational pattern, with visual cortex regions (yellow stars) clustering near early and middle vision model layers, while association cortices and limbic regions (blue and purple stars) positioned closer to language model representations. This visualization provides striking evidence for a hierarchical organization of representational alignment spanning both biological and artificial systems, with deeper processing stages across modalities converging toward similar representational strategies for high-level information integration.

\subsection*{Brain Alignment as a Precursor to Performance}

\begin{figure}[h]
\centering
\includegraphics[width=1.0\textwidth]{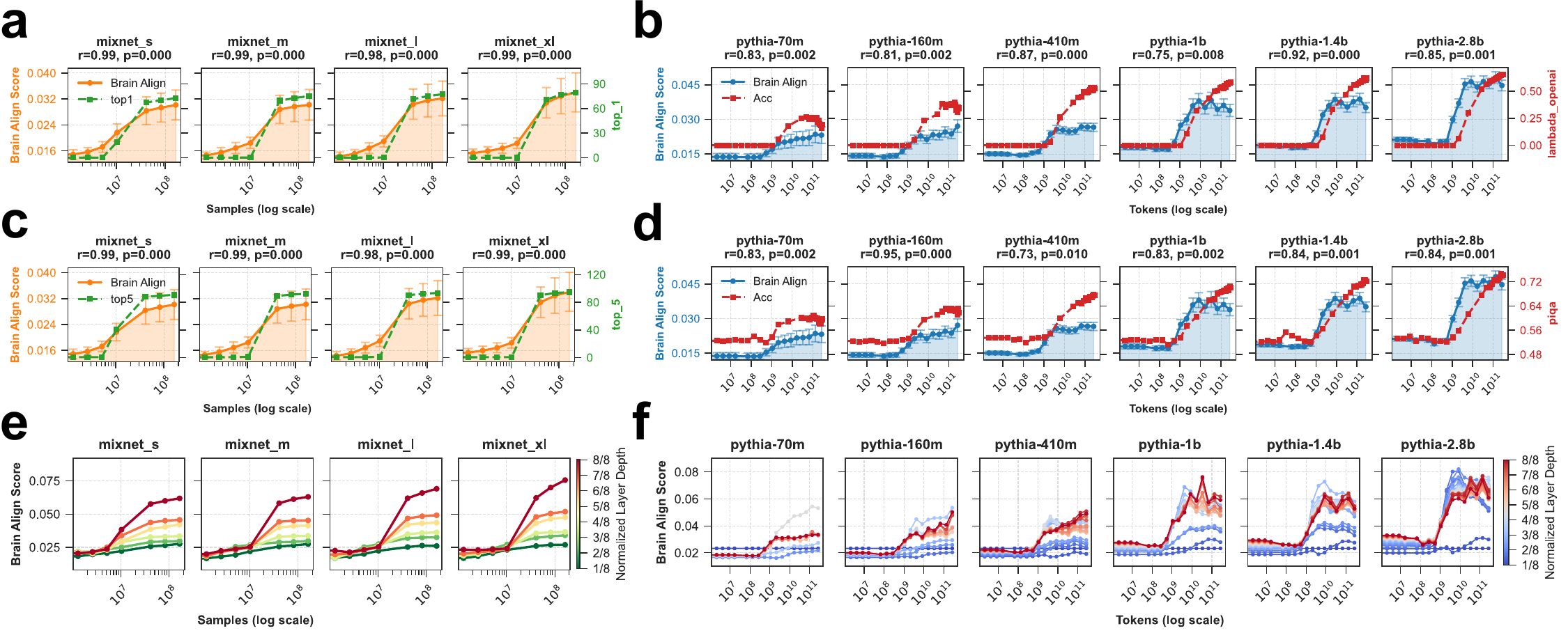}
\caption{\textbf{Evolution of brain alignment during model training.} \textbf{a-d}, Dual y-axis plots showing brain alignment score (left axis) and performance metrics (right axis) throughout training for vision models (MixNet family~\cite{tan2019mixconv}) and language models (Pythia family~\cite{biderman2023pythia}). \textbf{e-f}, Line plots showing layer-wise brain alignment evolution across training progression for vision and language models.}
\label{fig:training_evolution}
\end{figure} 

While our previous analyses established a robust correlation between model performance and brain alignment across different modalities and architectures, the temporal relationship between these variables remains unexplored. Understanding whether brain alignment precedes performance improvements—or vice versa—is crucial for distinguishing between two competing hypotheses: performance-driven convergence, where optimization for task success naturally leads to brain-like representations; and alignment-driven performance, where developing brain-like representations serves as a necessary stepping stone toward higher capabilities. Clarifying this temporal relationship could reveal whether the convergent evolution between artificial and biological systems follows a predictable developmental trajectory.

Although longitudinal neuroimaging studies in humans are constrained because development unfolds over years and data collection is resource-intensive, artificial neural networks provide a unique experimental paradigm. Saved checkpoints across training offer high-resolution temporal snapshots of representational evolution, thus enabling direct testing of whether brain alignment emerges as a precursor, a consequence, or a concurrent feature of performance improvements.

Longitudinal analysis of model training revealed that brain alignment is not merely a static property but evolves dynamically throughout the learning process (Figure~\ref{fig:training_evolution}a-d). For both vision models (MixNet family)~\cite{tan2019mixconv} and language models (Pythia family)~\cite{biderman2023pythia}, we tracked brain alignment alongside performance metrics (ImageNet~\cite{deng2009imagenet} top-1 and top-5 accuracy for vision models; LAMBADA~\cite{paperno2016lambada} and PIQA~\cite{bisk2020piqa} scores for language models) across training checkpoints.

A critical finding emerged from this analysis: increases in brain alignment consistently preceded improvements in task performance. For vision models, brain alignment increased throughout the entire training process, with particularly sharp rises in early training (reaching approximately 85\% of maximum alignment by just 20\% of training samples), while performance only began improving after approximately 5\% of training and continued to increase more gradually thereafter. The correlation between brain alignment and performance across the entire training process was strong ($r=0.98$ for MixNet-L). This pattern suggests that developing brain-like representations may be a necessary precursor to achieving high performance rather than merely a consequence of performance improvement.

This leading indicator relationship was even more pronounced in language models, where brain alignment began increasing as early as 1\% of training and rapidly reached stable levels by approximately 10\% of training tokens. In stark contrast, performance metrics on downstream tasks such as PIQA and LAMBADA showed significant delays, only reaching half of their maximum values after approximately 10\% of training. Across all Pythia models, we observed consistently strong correlations between brain alignment and task performance (ranging from $r=0.73$ to $r=0.95$, all $p<0.01$), with brain alignment increases consistently preceding performance improvements. This pattern held across model scales, reinforcing the hypothesis that developing brain-like representations may serve as a fundamental precursor to performance gains in language models.

The stabilization of overall brain alignment around 10\% of training is further elucidated by a layer-wise analysis, revealing distinct, modality-specific patterns (Figure~\ref{fig:training_evolution}e-f). In vision models, brain alignment steadily increased across all layers throughout training, with deeper layers ultimately achieving higher alignment scores than earlier layers.

In contrast, language models exhibited a different developmental trajectory. Early layers (normalized depth 1/8 to 2/8) maintained relatively stable and low alignment levels throughout training. This limited early-layer alignment can likely be attributed to the nature of the NSD dataset, which primarily features visual stimuli rather than rich linguistic content, offering little opportunity for alignment with token-level representations.

Middle layers (normalized depth 4/8 to 6/8) showed a rise-then-fall pattern: alignment initially increased during early training (up to 10\% of training tokens), followed by a moderate decline as training progressed. The deeper layers (normalized depth 7/8 to 8/8) exhibited a delayed growth pattern, with alignment continuing to increase even after middle layers had begun to decline, eventually stabilizing with minor fluctuations.

These layer-specific dynamics became increasingly pronounced in larger models. Models such as pythia-1b, pythia-1.4b, and pythia-2.8b exhibited more dramatic differentiation between early, middle, and deep layer behaviors compared to their smaller counterparts. This scale-dependent pattern suggests that larger models not only develop stronger brain alignment overall but also evolve more specialized and functionally differentiated internal representations across layers.

These training dynamics reinforce a central aspect of our convergent evolution hypothesis: as models become more effective at solving their target tasks, they naturally discover representational strategies that align with biological solutions, even in the absence of explicit constraints promoting brain-like processing. The observation that alignment precedes performance suggests that brain-like representations may be stepping stones toward effective task solutions rather than mere by-products of optimization.

\subsection*{Multi-scale Brain Alignment Reveals Regional Processing Hierarchies}

\begin{figure}[ht]
    \centering
    \includegraphics[width=1.0\textwidth]{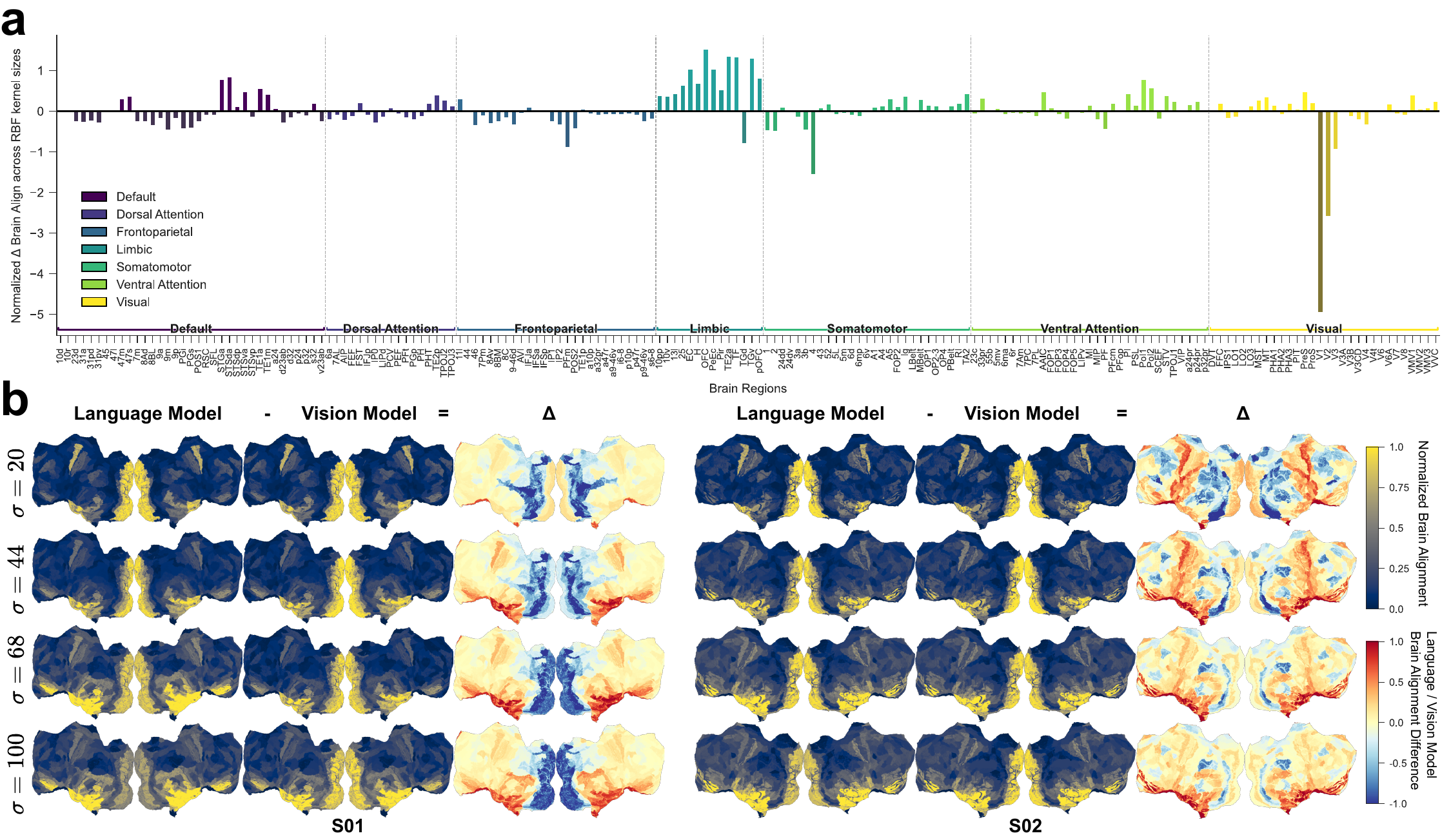}
    \caption{\textbf{Brain alignment patterns across representational scales.} \textbf{a}, Normalized brain alignment changes across different brain regions as kernel size increases from 28 to 68. Positive values indicate regions with increased relative alignment at larger representational scales, while negative values show decreased relative alignment. 
    \textbf{b}, Cortical surface visualizations showing alignment patterns between brain activity and both language models (left columns), vision models (middle columns), and their difference (right columns, $\Delta$) at four different kernel sizes. As kernel size increases, a posterior-to-anterior gradient emerges, with anterior regions showing relatively stronger alignment at larger scales, particularly for language models in frontal and temporal areas.}
    \label{fig:kernel_analysis}
\end{figure}

Previous research comparing brain activity patterns with AI representations has typically relied on single alignment metrics with fixed parameters, without systematically investigating how measurement scale impacts the observed correspondence between artificial and biological systems. While our analyses demonstrated significant consistency across different measures (Figure~\ref{fig:metric_consistency}), the choice of measurement scale may nonetheless reveal important organizational principles in both systems that would otherwise remain hidden. 

To explore this possibility, we systematically varied the kernel size in our CKA measurements from smaller to larger scales, uncovering a striking shift in which brain regions showed strongest model correspondence (Figure~\ref{fig:kernel_analysis}a). This methodological approach has direct biological significance: smaller kernel sizes (lower $\sigma$ values) focus on very similar activation patterns across stimuli, capturing highly specific representational similarities analogous to how primary sensory areas respond selectively to specific features. In contrast, larger kernel sizes (higher $\sigma$ values) detect similarities across more varied stimuli, reflecting broader representational patterns similar to how association cortices integrate diverse information across contexts.

Using the HCP\_MMP1~\cite{glasser2016multi} parcellation grouped according to the Yeo-7 network taxonomy~\cite{yeo2011organization}, we observed that as kernel size increased, primary visual areas showed substantial decreases in relative alignment, while limbic, ventral attention, and default mode networks showed increasing alignment. 

The most dramatic decreases in alignment with increasing kernel size were observed in early visual areas V1 ($\Delta = -5.30$), V2 ($\Delta = -2.81$), and V3 ($\Delta = -1.05$), while the strongest increases occurred in limbic regions including orbital frontal cortex (OFC, $\Delta = +1.56$), temporal regions TF ($\Delta = +1.36$), TE2a ($\Delta = +1.36$), and TGv ($\Delta = +1.33$), and parahippocampal regions EC ($\Delta = +1.09$) and PeEc ($\Delta = +1.06$).

These shifts were remarkably consistent across subjects, with a sign consistency of 0.83 in normalized alignment changes across all brain regions, suggesting a fundamental principle in how information is represented at different scales in the brain. 

This shift was visualized on the cortical surface in Figure~\ref{fig:kernel_analysis}b, revealing a gradient from posterior-to-anterior (occipital to frontal lobe) in alignment patterns at different kernel sizes. The cortical visualizations display the normalized brain alignment scores for both language and vision models across multiple kernel sizes. For language models, stronger alignment is visible in temporal and prefrontal regions compared to primary visual areas, particularly at larger kernel sizes. Vision models show similar patterns but with relatively stronger alignment in posterior visual areas. The difference maps ($\Delta$) highlight regions where language and vision models differ in their brain alignment, with red indicating stronger language model alignment and blue indicating stronger vision model alignment.

As kernel size increases, this posterior-to-anterior gradient becomes more pronounced across all subjects, with anterior regions showing relatively stronger alignment at larger scales. This differentiation is particularly evident for language models in frontal and temporal lobes at the largest kernel sizes. Visualizations for all four subjects and additional kernel size settings are provided in the supplementary materials.

These findings reveal a fundamental organizational principle shared by biological and artificial systems: as information flows through processing pathways, representations transition from local, feature-specific encodings—captured by smaller-scale analyses in primary sensory regions—to distributed, context-dependent structures better captured at larger scales in higher association areas. The stronger scale-dependent modulation observed in language models further suggests that language processing operates at inherently broader representational scales than vision, mirroring hierarchical information processing patterns observed in the brain.

\subsection*{Mapping the Landscape of AI Models via Brain Alignment}

\begin{figure}[h!]
    \centering
    \includegraphics[width=1.0\textwidth]{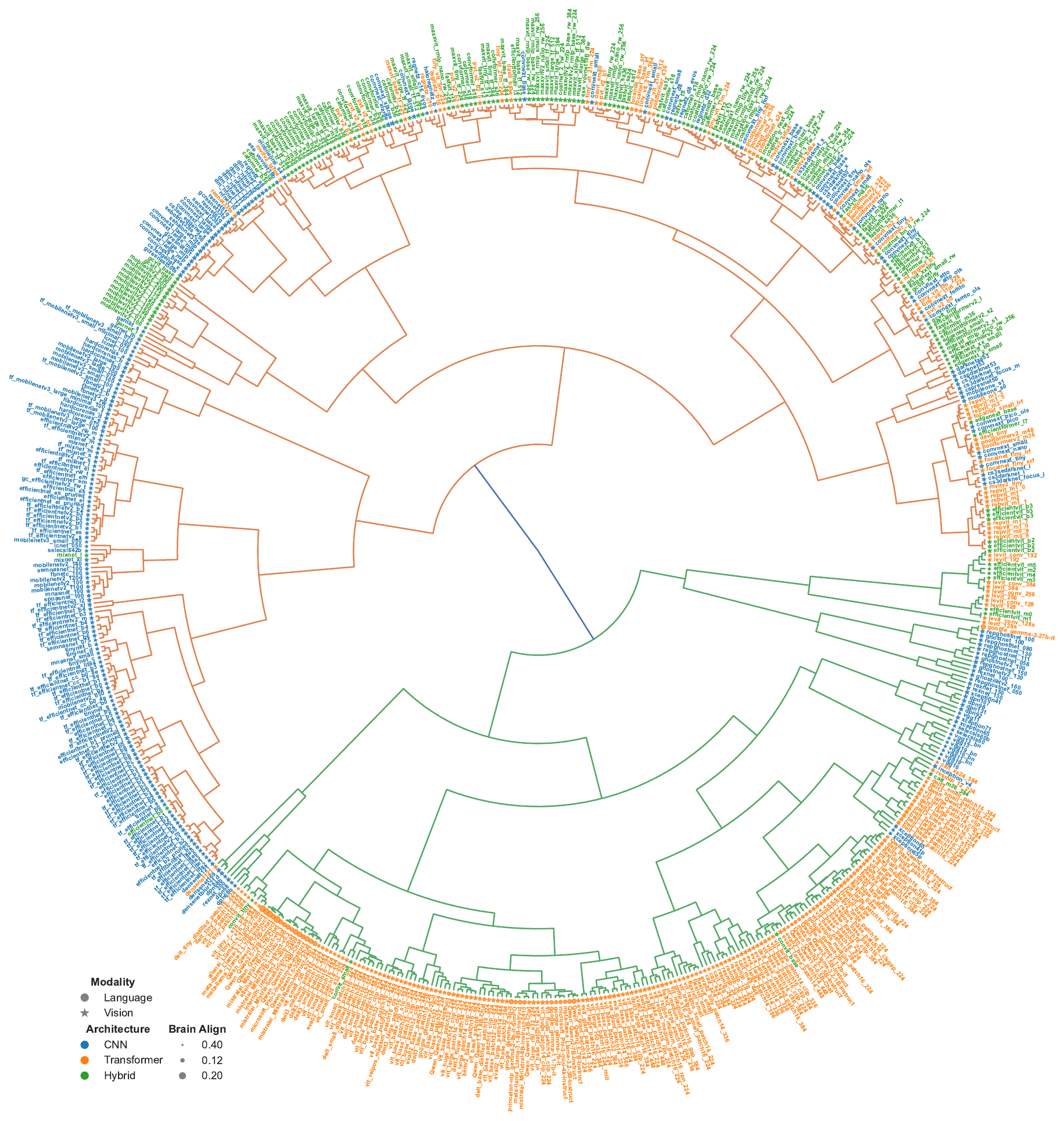}
    \caption{\textbf{Taxonomic organization of AI models based on brain alignment patterns.} Radial dendrogram showing hierarchical clustering of models based on their brain alignment vectors (180 brain regions $\times$ 8 normalized depth layers $\times$ 4 subjects). Each node represents a model, with node color indicating architecture class (CNN, Transformer, Hybrid), shape indicating modality (circles for language models, stars for vision models), and size representing the alignment score.}
    \label{fig:model_taxonomy}
\end{figure}

Finally, we examined how different models relate to each other when characterized by their brain alignment profiles. Using the full brain alignment pattern consisting of alignment scores across 180 cortical regions, 8 normalized depth levels, and all 4 subjects as a high-dimensional feature vector for each model, we constructed a hierarchical clustering tree to visualize representational similarity. The result is shown in Figure~\ref{fig:model_taxonomy}, where proximity reflects similarity in brain alignment patterns. 

Architecture emerges as the dominant organizing principle in the dendrogram. CNNs form two distinct clusters that reflect a generational and structural divide. One smaller cluster contains earlier CNN architectures such as VGG~\cite{simonyan2014very} and Xception~\cite{chollet2017xception}, which lack residual connections and rely on straightforward feedforward stacking of convolutional layers. In contrast, a larger and more internally coherent cluster includes modern CNNs like ResNet~\cite{he2016deep}, MobileNet~\cite{koonce2021mobilenetv3}, and EfficientNet~\cite{koonce2021efficientnet}, which employ residual or shortcut connections. These architectural innovations appear to yield more brain-aligned representations. The tight grouping of modern CNNs suggests that residual structures impose a stronger inductive bias toward biologically plausible representations, while the earlier models diverge both in form and function.

Transformer-based models occupy a broader and more internally diverse region of the tree. Within this space, language models form several well-defined subclusters that often correspond to model families such as LLaMA~\cite{grattafiori2024llama} or Qwen~\cite{yang2024qwen2}, suggesting that architectural lineage and fine-tuning methodology significantly shape alignment patterns. Language models tend to cluster separately from vision models, even among transformers, indicating a modality-driven divergence in representational geometry at the model level.

Hybrid architectures, which integrate convolutional and attention-based components (e.g., ConvNeXt~\cite{liu2022convnet}, CoAtNet~\cite{dai2021coatnet}, FocalNet~\cite{yang2022focal}), display a more scattered distribution across the dendrogram. Hybrid architectures display diverse alignment profiles, distributing across the taxonomic space based on their relative balance of convolutional and attentional components. This dispersion reflects their architectural flexibility and suggests that hybrid models do not converge toward a single representational style. Instead, they form a continuum of brain alignment profiles.

This alignment-driven taxonomy offers a biologically grounded organizational view of the model landscape. It complements traditional task benchmarks by revealing deeper commonalities and distinctions in representational structure. These results reinforce the broader theme of our study: representational convergence between artificial and biological systems is structured and quantifiable, and emerges in systematic ways across model families, architectures, and modalities.




 

\section*{Discussion}

Our findings provide compelling evidence for convergent evolution between artificial and biological intelligence systems. Despite fundamentally different origins, architectures, and learning mechanisms, AI models spontaneously develop representations that progressively align with human brain activity patterns as they optimize for task performance. This emergent alignment occurs without explicit neurobiological constraints, suggesting that certain computational solutions may be universal for intelligent information processing, transcending specific physical implementations~\cite{hasson2020direct, richards2019deep, zador2019critique}.

\paragraph{Performance Alignment and Convergent Evolution} The robust correlation between model performance and brain alignment across both language and vision domains represents perhaps the most striking evidence for convergent evolution. The logarithmic relationship observed in both modalities suggests a principled trajectory of diminishing returns, where initial improvements yield substantial alignment gains followed by more modest increases. This pattern aligns with theoretical perspectives suggesting that as complex systems approach optimal solutions for information processing challenges, they naturally converge toward similar computational strategies despite different evolutionary pathways~\cite{yamins2014performance, marblestone2016toward}. The substantially stronger correlation observed in language models compared to vision models ($r=0.89$ vs. $r=0.53$) suggests that language processing may impose more stringent computational constraints that channel solutions toward brain-like implementations~\cite{schrimpf2021neural, caucheteux2022brains}. This modality difference may reflect the heightened abstraction and contextual integration demands of language compared to visual perception~\cite{fedorenko2014reworking, pereira2018toward}.


\paragraph{Hierarchical and Multi-scale Organization} Layer-wise and multi-scale alignment analyses revealed systematic gradients of representational organization. Vision models show a gradual increase in brain alignment with depth, paralleling the hierarchical progression from low-level to high-level visual processing in the brain~\cite{yamins2016using, cichy2016comparison}. Language models exhibit stronger and more distributed alignment, corresponding to association areas involved in abstract semantic processing~\cite{goldstein2022shared, dobs2022brain}. Importantly, deeper vision model layers begin to resemble language models, suggesting cross-modal convergence at higher abstraction levels~\cite{binder2016toward, huth2016natural}. Multi-scale analysis further revealed a posterior-to-anterior shift in alignment across brain regions, consistent with a transition from localized, feature-specific representations to distributed, integrative ones~\cite{margulies2016situating, huntenburg2018large}. The stronger modulation in language models supports the notion that language operates on inherently broader representational scales~\cite{fedorenko2016language, hagoort2019neurobiology}.

\paragraph{Temporal Dynamics and Architectural Inductive Biases}
Longitudinal analysis revealed that brain alignment consistently precedes improvements in task performance, suggesting that brain-like representations may function as computational stepping stones rather than mere by-products of training~\cite{saxe2019mathematical, kriegeskorte2015deep}. This phenomenon was observed across architectures and scales, reinforcing the idea that convergent evolution arises as models optimize for performance. Taxonomic clustering further showed that architectural inductive biases strongly shape alignment profiles. CNNs and transformers formed distinct clusters, while hybrid models exhibited a continuum of profiles. Models with residual connections and attention mechanisms exhibited higher brain alignment scores in our analysis, suggesting these architectural elements may promote representational structures that more closely resemble biological systems~\cite{kriegeskorte2018cognitive, kietzmann2019recurrence, kubilius2019brain}.

\paragraph{Limitations and Implications}
Our study has several limitations. First, we rely on static fMRI data with limited temporal resolution, which may overlook fine-grained dynamics. Second, the NSD dataset centers on passive visual processing, limiting insight into language-specific or abstract reasoning processes. Third, benchmark contamination may inflate performance estimates for large models trained on internet-scale corpora~\cite{dodge2021documenting, bender2021dangers}. Future studies could integrate richer behavioral datasets and temporally resolved neural recordings to address these issues.

These findings have broad implications. For neuroscience, they suggest that AI models optimized for general tasks may spontaneously recapitulate brain-like representations, supporting the idea that cortical organization reflects efficient computational solutions~\cite{kriegeskorte2019interpreting, richards2019deep}. For AI, they imply that brain-aligned architectures or objectives could guide model development and improve interpretability~\cite{hassabis2017neuroscience, lindsay2021convolutional}. Ultimately, understanding how and why alignment emerges may bridge the gap between natural and artificial intelligence.

\section*{Methods}


To systematically investigate brain-AI alignment, we developed a comprehensive analytical framework combining neuroimaging and machine learning techniques. This approach quantifies similarities between neural activity patterns in human brains and internal representations in AI models using Centered Kernel Alignment (CKA)~\cite{kornblith2019similarity}, a robust similarity metric that accounts for different dimensionalities across systems. We analyzed fMRI responses from human subjects viewing natural images (Natural Scenes Dataset)~\cite{allen2022massive} alongside activations from 630 AI models (spanning vision and language domains) processing the same stimuli. By varying measurement parameters and examining alignment across different brain regions, model architectures, and training stages, we uncovered systematic patterns of representational convergence between biological and artificial systems. Statistical robustness was ensured through extensive validation across multiple subjects, metrics, and analytical parameters.

\subsection*{Data Acquisition and Preprocessing}
\paragraph{fMRI Data}
For our analysis, we utilized the Natural Scenes Dataset (NSD)~\cite{allen2022massive}, a large-scale fMRI dataset capturing neural responses to natural images. From the eight subjects in the dataset, we selected four (S01, S02, S05, S07) who had complete data trajectories. Each subject's data included responses to 24,980 training stimuli (unique per subject) and 2,770 testing stimuli (shared across subjects). To maximize data volume while ensuring consistent analysis conditions, we focused on the training set for our primary analyses.

We extracted functional responses from the preprocessed \texttt{func1pt8mm} data following the standard NSD preprocessing pipeline, which includes motion correction, distortion correction, and hemodynamic response modeling to extract beta weights representing neural responses to each stimulus. All analyses were conducted in subject-specific functional space to preserve individual neural response patterns.

\paragraph{Stimulus Materials}
The NSD stimuli consist of images from the COCO dataset~\cite{lin2014microsoft}, which provides paired image-caption annotations. For each image, we extracted the first available caption to serve as the textual input for language models. This approach enabled aligned multimodal analyses across vision and language domains using identical stimuli, allowing for direct comparison of neural representations across modalities.

\subsection*{Neural Network Models and Representation Extraction}

\paragraph{Language Models}
We analyzed 36 contemporary language models spanning diverse architectures and parameter scales (500M to 72B parameters). Our selection included major model families: Qwen~\cite{yang2024qwen2}, Llama~\cite{grattafiori2024llama}, Gemma~\cite{team2024gemma}, Mistral~\cite{jiangMistral7B2023}, and Phi~\cite{abdin2024phi}, among others~\cite{team2023internlm}. To explore the effects of different fine-tuning approaches, we included specialized variants such as SimPO~\cite{meng2024simpo} and Tulu~\cite{lambert2024t}.

For consistent representation extraction, we implemented all models using the HuggingFace Transformers library~\cite{wolf-etal-2020-transformers} with default configurations. Each model processed image captions using its standard chat template. For example, with Llama-3 models, the template followed this structure:

\begin{verbatim}
<|start_header_id|>user<|end_header_id|>
Consider the image described by the following sentences: 
    '{caption}' 
<|start_header_id|>assistant<|end_header_id|>
\end{verbatim}

From each model, we extracted hidden state representations from the output of each transformer block, focusing on the last token representation, which is a standard approach for capturing sentence-level semantics~\cite{devlin2019bert, liu2019roberta, bommasani2021opportunities}. For performance evaluation, we used LLM Leaderboard 2~\cite{open-llm-leaderboard-v2}, a composite benchmark aggregating performance across multiple tasks including BBH (Big-Bench Hard)~\cite{suzgun2023challenging}, MMLU-PRO~\cite{wang2024mmlu}, IFEval~\cite{zhou2023instruction}, GPQA~\cite{rein2024gpqa}, MATH~\cite{hendrycks2measuring}, and MUSR~\cite{sprague2024musr}, covering various reasoning, knowledge, and language understanding dimensions. We also incorporated Chatbot Arena rankings~\cite{chiang2024chatbot}~\cite{chiang2024chatbot}, which offers complementary human evaluations through a pairwise preference system rather than automated metrics.

\paragraph{Vision Models}
We analyzed 594 vision models spanning three architectural categories: Convolutional Neural Networks (CNNs), Transformers, and Hybrid models. All models were sourced from the TIMM library~\cite{rw2019timm} with pre-trained weights, encompassing parameter ranges from 1.33M to 1.014B. Due to the heterogeneous structure of vision architectures, we implemented an automated feature extraction pipeline to capture hidden states from each network block across all models, extracting outputs from each individual feature extraction component throughout the network hierarchy.

For each image stimulus, we applied the standard preprocessing procedure specified for each model (typically normalizing according to ImageNet statistics). Model performance was evaluated using multiple benchmarks, including ImageNet-1K top-1 accuracy~\cite{deng2009imagenet}, ImageNet-A~\cite{hendrycks2021natural}, ImageNet-R~\cite{hendrycks2021many}, and other variant datasets as detailed in Table~\ref{tab:correlation_stats}.

\subsection*{Brain Parcellation and Anatomical Framework}

We employed the HCP\_MMP1 parcellation~\cite{glasser2016multi}, which divides the cortex into 180 anatomically and functionally distinct regions. To analyze functional network properties, we mapped these regions to the Yeo-7 network taxonomy~\cite{yeo2011organization}, which groups cortical areas into seven functional networks: Visual, Somatomotor, Dorsal Attention, Ventral Attention, Limbic, Frontoparietal, and Default. This mapping followed the correspondence established by Byrge and Kennedy~\cite{byrge2019high}, enabling analysis of representational similarities at both region-specific and network levels.

\subsection*{Representational Similarity Analysis}

\paragraph{Centered Kernel Alignment}

To quantify representational similarity between AI model activations and brain activity patterns, we employed CKA, a robust similarity index that accounts for different dimensionalities and is invariant to orthogonal transformations. The CKA between two representation matrices $X$ and $Y$ is defined as:
\begin{equation}
    \text{CKA}(K, L) = \frac{\text{HSIC}(K, L)}{\sqrt{\text{HSIC}(K, K) \times \text{HSIC}(L, L)}},
\end{equation}
where $K$ and $L$ are kernel matrices derived from $X$ and $Y$, and HSIC is the Hilbert-Schmidt Independence Criterion. For our analyses, we utilized the Radial Basis Function (RBF) kernel:
\begin{equation}
    K(x_i, x_j) = \exp\left(-\frac{\|x_i - x_j\|^2}{2\sigma^2}\right),
\end{equation}
where $\sigma$ controls the kernel width. For computational efficiency with high-dimensional representations exceeding 4096 dimensions, we applied Principal Component Analysis (PCA) to reduce dimensionality while preserving at least 95\% of variance. All CKA computations were implemented in PyTorch and executed on GPUs with parallelization across models and brain regions.

We performed analyses across multiple kernel sizes (20, 28, 36, 44, 52, 60, 68, 72, 84, 92, 100) to capture representational similarities at different scales, from local feature-specific patterns to global representational structures. This multi-scale approach enabled detection of alignment patterns that might be specific to particular representational granularities. Despite observing scale-dependent patterns as reported in the Results section, we found high consistency in alignment patterns across different kernel sizes, confirming the robustness of our observations.

\begin{figure}[ht]
    \centering
    \includegraphics[width=\textwidth]{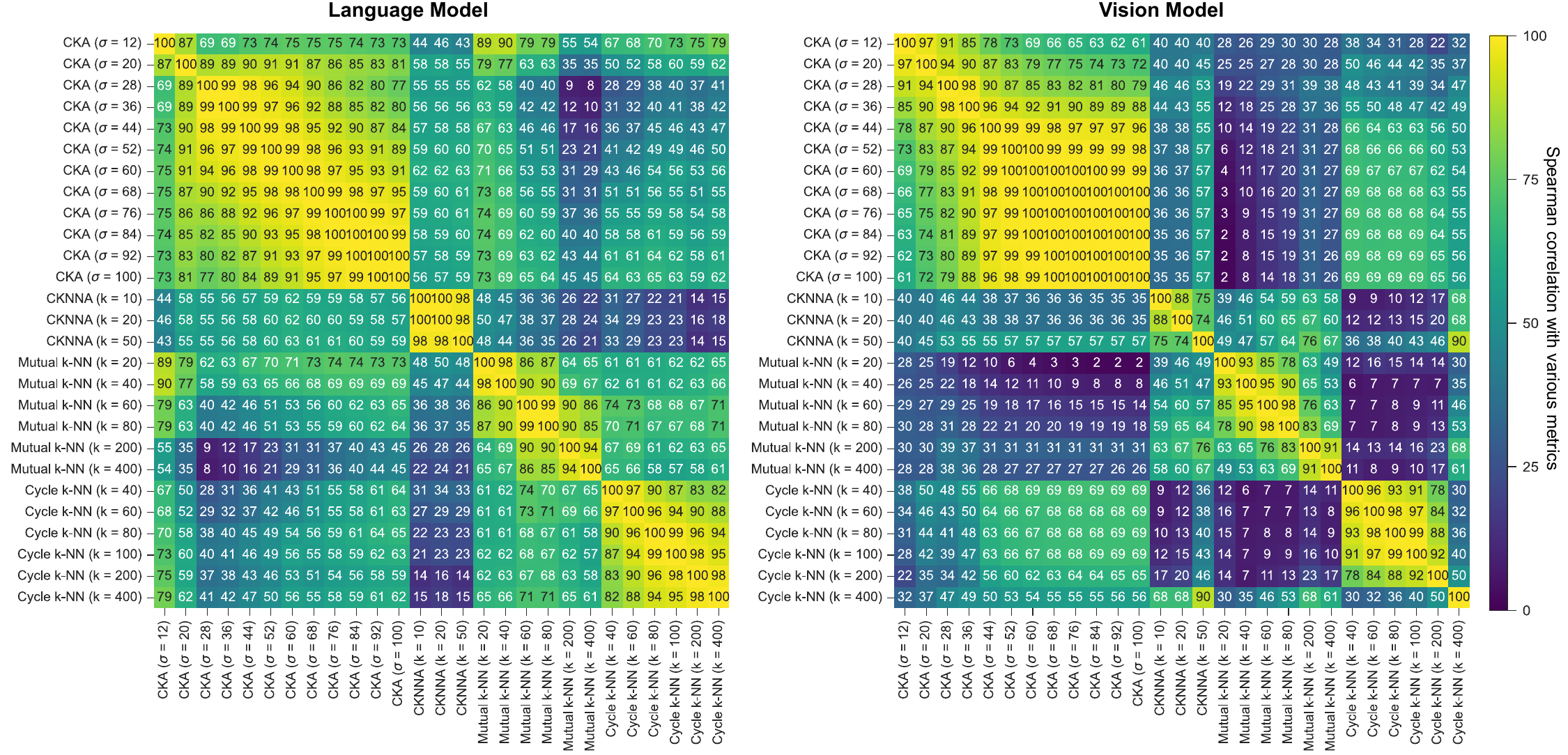}
    \caption{\textbf{Consistency analysis across different similarity metrics and parameter settings.} Heatmaps showing Spearman rank correlations between different similarity measurement approaches for language models and vision models. The analysis compares CKA with varying kernel sizes, k-nearest neighbor (KNN) variants with different $k$ values. The consistently high correlation values indicate strong agreement in how different metrics rank the similarity relationships, demonstrating the robustness of our representational alignment framework across methodological variations.}
    \label{fig:metric_consistency}
\end{figure}

To further validate the stability of our similarity metrics, we conducted a comprehensive consistency analysis across different kernel sizes and alternative k-nearest neighbor (KNN) based similarity measures (Figure~\ref{fig:metric_consistency}). Using Spearman rank correlation, we evaluated whether the relative rankings of model similarities remained consistent despite variations in the similarity computation method. This analysis confirmed strong rank consistency across different parameter settings, supporting the reliability of our representational similarity framework regardless of specific kernel size choices.

For brain-region specific analyses, we calculated alignment scores for each model by taking the average of the top 5\% highest-aligning brain regions at the normalized model depth of 6/8. This approach highlighted the brain regions most strongly corresponding to model representations, providing a more sensitive measure of alignment than averaging across all regions.

\subsection*{Longitudinal Training Analysis}

To investigate how brain alignment evolves during model training, we leveraged two model families with available training checkpoints:

\begin{itemize}
    \item For language models, we analyzed the Pythia family~\cite{biderman2023pythia}, which provides numerous checkpoints throughout training for models ranging from 70M to 2.8B parameters. These checkpoints enabled tracking of representational evolution across training trajectories.
    \item For vision models, we utilized the MixNet family~\cite{tan2019mixconv}, training multiple variants (S/M/L/XL) from scratch on ImageNet-1K~\cite{krizhevsky2012imagenet} using standard TIMM~\cite{rw2019timm} configurations. Training checkpoints were saved at logarithmically spaced intervals to capture both early and late training dynamics.
\end{itemize}

For language models, we tracked performance on LAMBADA~\cite{paperno2016lambada} (language modeling) and PIQA~\cite{bisk2020piqa} (physical common sense reasoning) benchmarks across training checkpoints. For vision models, we monitored ImageNet top-1 and top-5 accuracy. At each checkpoint, we calculated brain alignment scores using the CKA methodology described above, enabling temporal correlation analysis between brain alignment and task performance. Pearson correlation coefficients were computed to quantify the relationship between alignment and performance trajectories throughout training.

\subsection*{Multi-scale Analysis Methods}

\paragraph{Regional Sensitivity to Kernel Size}

To analyze how different brain regions respond to variations in representational scale, we compared alignment patterns across different kernel sizes. For each brain region, we calculated normalized alignment scores by first averaging across all models and then computing the difference ($\Delta$) in alignment between larger kernel sizes (68) and smaller kernel sizes (28). Positive $\Delta$ values indicate regions with relatively stronger alignment at larger scales (capturing more global representational properties), while negative values indicate regions with stronger alignment at smaller scales (capturing more local features).

To validate the consistency of these patterns, we calculated the sign consistency of normalized alignment changes across subjects, finding a high consistency value of 0.83, which confirms the robustness of the observed scale-dependent regional preferences across individuals.

\subsection*{Model Taxonomy Construction}

We constructed a hierarchical organization of models based on their brain alignment profiles using agglomerative clustering. For each model, we created a feature vector comprising alignment scores across all 180 brain regions and 8 normalized depth levels for all 4 subjects, resulting in a 180 $\times$ 8 $\times$ 4 dimensional representation. We computed pairwise distances using cosine similarity and applied Ward's linkage method~\cite{ward1963hierarchical} to generate the hierarchical clustering structure. The resulting dendrogram was visualized as a radial tree diagram with node attributes (size, color, shape) representing brain alignment score, architectural class, and modality, respectively.

\subsection*{Statistical Analysis Framework}

\paragraph{Correlation Analysis}
We assessed the relationship between model performance and brain alignment using both Pearson and Spearman correlation coefficients, capturing linear and monotonic relationships respectively. For each correlation, we calculated 95\% confidence intervals via bootstrap sampling with 1,000 iterations. To account for multiple comparisons across different performance metrics and model types, we applied False Discovery Rate (FDR) correction to p-values.

\paragraph{Regression Models}
To characterize the relationship between model performance and brain alignment, we fitted both linear and logarithmic regression models. Model selection was based on Akaike Information Criterion (AIC), Bayesian Information Criterion (BIC), and $R^2$ values. For the language model performance-alignment relationship, a logarithmic model ($R^2 = 0.80$, AIC = 117.6, BIC = 120.8) outperformed a linear model ($R^2 = 0.78$, AIC = 120.7, BIC = 123.9), indicating a diminishing returns pattern. Similarly, for vision models, a logarithmic fit ($R^2 = 0.33$, AIC = -3810.2, BIC = -3801.4) provided better characterization than a linear model ($R^2 = 0.28$, AIC = -3772.1, BIC = -3763.3).

\paragraph{Inter-subject Consistency Analysis}
\label{sec:inter_subject}
\begin{table}[h]
    \centering
    \begin{minipage}{0.48\textwidth}
        \centering
        \caption{\textbf{Inter-subject consistency for language models.} Pearson correlation ($r$) values between brain alignment patterns across subjects for language models.}
        \label{tab:lang_corr}
        \begin{tabular}{l|cccc}
            \hline
                & S01   & S02   & S05   & S07 \\
            \hline
            S01 & 1.000 & 0.997 & 0.999 & 0.997 \\
            S02 & 0.997 & 1.000 & 0.999 & 1.000 \\
            S05 & 0.999 & 0.999 & 1.000 & 0.999 \\
            S07 & 0.997 & 1.000 & 0.999 & 1.000 \\
            \hline
        \end{tabular}
    \end{minipage}
    \hfill 
    \begin{minipage}{0.48\textwidth}
        \centering
        \caption{\textbf{Inter-subject consistency for vision models.} Pearson correlation ($r$) values between brain alignment patterns across subjects for vision models.}
        \label{tab:vision_corr}
        \begin{tabular}{l|cccc}
            \hline
                & S01   & S02   & S05   & S07 \\
            \hline
            S01 & 1.000 & 0.970 & 0.981 & 0.970 \\
            S02 & 0.970 & 1.000 & 0.998 & 0.999 \\
            S05 & 0.980 & 0.998 & 1.000 & 0.998 \\
            S07 & 0.970 & 0.999 & 0.998 & 1.00 \\
            \hline
        \end{tabular}
    \end{minipage}
\end{table}

To ensure the robustness of our findings, we conducted a comprehensive analysis to determine whether the relationship between brain alignment scores and model performance was consistent across different individuals. We calculated pairwise Pearson correlations between these brain-performance relationships for all four subjects (S01, S02, S05, S07), as shown in Tables~\ref{tab:lang_corr} and~\ref{tab:vision_corr}. These tables demonstrate the consistency of how brain alignment relates to model performance across different individuals.

For language models, we observed remarkably high inter-subject correlations, with values ranging from 0.997 to 1.000. This near-perfect consistency indicates that the pattern of which brain regions align with language model representations is highly preserved across different individuals. For vision models, we also found extremely strong consistency, with correlations ranging from 0.970 to 0.999. 

These extraordinarily high inter-subject correlations for both modalities provide compelling evidence that the regional patterns of brain-AI alignment we observed reflect fundamental organizational principles of neural information processing universally shared across human brains rather than idiosyncratic individual variations.

\bibliography{sample}

\begin{thebibliography}{10}
\urlstyle{rm}
\expandafter\ifx\csname url\endcsname\relax
  \def\url#1{\texttt{#1}}\fi
\expandafter\ifx\csname urlprefix\endcsname\relax\def\urlprefix{URL }\fi
\expandafter\ifx\csname doiprefix\endcsname\relax\def\doiprefix{DOI: }\fi
\providecommand{\bibinfo}[2]{#2}
\providecommand{\eprint}[2][]{\url{#2}}

\bibitem{lecun2015deep}
\bibinfo{author}{LeCun, Y.}, \bibinfo{author}{Bengio, Y.} \& \bibinfo{author}{Hinton, G.}
\newblock \bibinfo{journal}{\bibinfo{title}{Deep learning}}.
\newblock {\emph{\JournalTitle{nature}}} \textbf{\bibinfo{volume}{521}}, \bibinfo{pages}{436--444} (\bibinfo{year}{2015}).

\bibitem{silver2016mastering}
\bibinfo{author}{Silver, D.} \emph{et~al.}
\newblock \bibinfo{journal}{\bibinfo{title}{Mastering the game of go with deep neural networks and tree search}}.
\newblock {\emph{\JournalTitle{nature}}} \textbf{\bibinfo{volume}{529}}, \bibinfo{pages}{484--489} (\bibinfo{year}{2016}).

\bibitem{brown2020language}
\bibinfo{author}{Brown, T.} \emph{et~al.}
\newblock \bibinfo{journal}{\bibinfo{title}{Language models are few-shot learners}}.
\newblock {\emph{\JournalTitle{Advances in neural information processing systems}}} \textbf{\bibinfo{volume}{33}}, \bibinfo{pages}{1877--1901} (\bibinfo{year}{2020}).

\bibitem{jones2025large}
\bibinfo{author}{Jones, C.~R.} \& \bibinfo{author}{Bergen, B.~K.}
\newblock \bibinfo{journal}{\bibinfo{title}{Large language models pass the turing test}}.
\newblock {\emph{\JournalTitle{arXiv preprint arXiv:2503.23674}}}  (\bibinfo{year}{2025}).

\bibitem{grattafiori2024llama}
\bibinfo{author}{Grattafiori, A.} \emph{et~al.}
\newblock \bibinfo{journal}{\bibinfo{title}{The llama 3 herd of models}}.
\newblock {\emph{\JournalTitle{arXiv preprint arXiv:2407.21783}}}  (\bibinfo{year}{2024}).

\bibitem{yang2024qwen2}
\bibinfo{author}{Yang, A.} \emph{et~al.}
\newblock \bibinfo{journal}{\bibinfo{title}{Qwen2. 5 technical report}}.
\newblock {\emph{\JournalTitle{arXiv preprint arXiv:2412.15115}}}  (\bibinfo{year}{2024}).

\bibitem{dosovitskiy2020image}
\bibinfo{author}{Dosovitskiy, A.} \emph{et~al.}
\newblock \bibinfo{journal}{\bibinfo{title}{An image is worth 16x16 words: Transformers for image recognition at scale}}.
\newblock {\emph{\JournalTitle{arXiv preprint arXiv:2010.11929}}}  (\bibinfo{year}{2020}).

\bibitem{bommasani2021opportunities}
\bibinfo{author}{Bommasani, R.} \emph{et~al.}
\newblock \bibinfo{journal}{\bibinfo{title}{On the opportunities and risks of foundation models}}.
\newblock {\emph{\JournalTitle{arXiv preprint arXiv:2108.07258}}}  (\bibinfo{year}{2021}).

\bibitem{yamins2014performance}
\bibinfo{author}{Yamins, D.~L.} \emph{et~al.}
\newblock \bibinfo{journal}{\bibinfo{title}{Performance-optimized hierarchical models predict neural responses in higher visual cortex}}.
\newblock {\emph{\JournalTitle{Proceedings of the national academy of sciences}}} \textbf{\bibinfo{volume}{111}}, \bibinfo{pages}{8619--8624} (\bibinfo{year}{2014}).

\bibitem{schrimpf2021neural}
\bibinfo{author}{Schrimpf, M.} \emph{et~al.}
\newblock \bibinfo{journal}{\bibinfo{title}{The neural architecture of language: Integrative modeling converges on predictive processing}}.
\newblock {\emph{\JournalTitle{Proceedings of the National Academy of Sciences}}} \textbf{\bibinfo{volume}{118}}, \bibinfo{pages}{e2105646118} (\bibinfo{year}{2021}).

\bibitem{storrs2021diverse}
\bibinfo{author}{Storrs, K.~R.}, \bibinfo{author}{Kietzmann, T.~C.}, \bibinfo{author}{Walther, A.}, \bibinfo{author}{Mehrer, J.} \& \bibinfo{author}{Kriegeskorte, N.}
\newblock \bibinfo{journal}{\bibinfo{title}{Diverse deep neural networks all predict human inferior temporal cortex well, after training and fitting}}.
\newblock {\emph{\JournalTitle{Journal of cognitive neuroscience}}} \textbf{\bibinfo{volume}{33}}, \bibinfo{pages}{2044--2064} (\bibinfo{year}{2021}).

\bibitem{zhuang2017toward}
\bibinfo{author}{Zhuang, C.}, \bibinfo{author}{Kubilius, J.}, \bibinfo{author}{Hartmann, M.~J.} \& \bibinfo{author}{Yamins, D.~L.}
\newblock \bibinfo{journal}{\bibinfo{title}{Toward goal-driven neural network models for the rodent whisker-trigeminal system}}.
\newblock {\emph{\JournalTitle{Advances in Neural Information Processing Systems}}} \textbf{\bibinfo{volume}{30}} (\bibinfo{year}{2017}).

\bibitem{yamins2016using}
\bibinfo{author}{Yamins, D.~L.} \& \bibinfo{author}{DiCarlo, J.~J.}
\newblock \bibinfo{journal}{\bibinfo{title}{Using goal-driven deep learning models to understand sensory cortex}}.
\newblock {\emph{\JournalTitle{Nature neuroscience}}} \textbf{\bibinfo{volume}{19}}, \bibinfo{pages}{356--365} (\bibinfo{year}{2016}).

\bibitem{nilsson2013eye}
\bibinfo{author}{Nilsson, D.-E.}
\newblock \bibinfo{journal}{\bibinfo{title}{Eye evolution and its functional basis}}.
\newblock {\emph{\JournalTitle{Visual neuroscience}}} \textbf{\bibinfo{volume}{30}}, \bibinfo{pages}{5--20} (\bibinfo{year}{2013}).

\bibitem{ogura2004comparative}
\bibinfo{author}{Ogura, A.}, \bibinfo{author}{Ikeo, K.} \& \bibinfo{author}{Gojobori, T.}
\newblock \bibinfo{journal}{\bibinfo{title}{Comparative analysis of gene expression for convergent evolution of camera eye between octopus and human}}.
\newblock {\emph{\JournalTitle{Genome research}}} \textbf{\bibinfo{volume}{14}}, \bibinfo{pages}{1555--1561} (\bibinfo{year}{2004}).

\bibitem{zador2019critique}
\bibinfo{author}{Zador, A.~M.}
\newblock \bibinfo{journal}{\bibinfo{title}{A critique of pure learning and what artificial neural networks can learn from animal brains}}.
\newblock {\emph{\JournalTitle{Nature communications}}} \textbf{\bibinfo{volume}{10}}, \bibinfo{pages}{3770} (\bibinfo{year}{2019}).

\bibitem{lillicrap2020backpropagation}
\bibinfo{author}{Lillicrap, T.~P.}, \bibinfo{author}{Santoro, A.}, \bibinfo{author}{Marris, L.}, \bibinfo{author}{Akerman, C.~J.} \& \bibinfo{author}{Hinton, G.}
\newblock \bibinfo{journal}{\bibinfo{title}{Backpropagation and the brain}}.
\newblock {\emph{\JournalTitle{Nature Reviews Neuroscience}}} \textbf{\bibinfo{volume}{21}}, \bibinfo{pages}{335--346} (\bibinfo{year}{2020}).

\bibitem{hassabis2017neuroscience}
\bibinfo{author}{Hassabis, D.}, \bibinfo{author}{Kumaran, D.}, \bibinfo{author}{Summerfield, C.} \& \bibinfo{author}{Botvinick, M.}
\newblock \bibinfo{journal}{\bibinfo{title}{Neuroscience-inspired artificial intelligence}}.
\newblock {\emph{\JournalTitle{Neuron}}} \textbf{\bibinfo{volume}{95}}, \bibinfo{pages}{245--258} (\bibinfo{year}{2017}).

\bibitem{richards2019deep}
\bibinfo{author}{Richards, B.~A.} \emph{et~al.}
\newblock \bibinfo{journal}{\bibinfo{title}{A deep learning framework for neuroscience}}.
\newblock {\emph{\JournalTitle{Nature neuroscience}}} \textbf{\bibinfo{volume}{22}}, \bibinfo{pages}{1761--1770} (\bibinfo{year}{2019}).

\bibitem{allen2022massive}
\bibinfo{author}{Allen, E.~J.} \emph{et~al.}
\newblock \bibinfo{journal}{\bibinfo{title}{A massive 7t fmri dataset to bridge cognitive neuroscience and artificial intelligence}}.
\newblock {\emph{\JournalTitle{Nature neuroscience}}} \textbf{\bibinfo{volume}{25}}, \bibinfo{pages}{116--126} (\bibinfo{year}{2022}).

\bibitem{lin2014microsoft}
\bibinfo{author}{Lin, T.-Y.} \emph{et~al.}
\newblock \bibinfo{title}{Microsoft coco: Common objects in context}.
\newblock In \emph{\bibinfo{booktitle}{Computer vision--ECCV 2014: 13th European conference, zurich, Switzerland, September 6-12, 2014, proceedings, part v 13}}, \bibinfo{pages}{740--755} (\bibinfo{organization}{Springer}, \bibinfo{year}{2014}).

\bibitem{team2024gemma}
\bibinfo{author}{Team, G.} \emph{et~al.}
\newblock \bibinfo{journal}{\bibinfo{title}{Gemma 2: Improving open language models at a practical size}}.
\newblock {\emph{\JournalTitle{arXiv preprint arXiv:2408.00118}}}  (\bibinfo{year}{2024}).

\bibitem{krizhevsky2012imagenet}
\bibinfo{author}{Krizhevsky, A.}, \bibinfo{author}{Sutskever, I.} \& \bibinfo{author}{Hinton, G.~E.}
\newblock \bibinfo{journal}{\bibinfo{title}{Imagenet classification with deep convolutional neural networks}}.
\newblock {\emph{\JournalTitle{Advances in neural information processing systems}}} \textbf{\bibinfo{volume}{25}} (\bibinfo{year}{2012}).

\bibitem{kornblith2019similarity}
\bibinfo{author}{Kornblith, S.}, \bibinfo{author}{Norouzi, M.}, \bibinfo{author}{Lee, H.} \& \bibinfo{author}{Hinton, G.}
\newblock \bibinfo{title}{Similarity of neural network representations revisited}.
\newblock In \emph{\bibinfo{booktitle}{International conference on machine learning}}, \bibinfo{pages}{3519--3529} (\bibinfo{organization}{PMLR}, \bibinfo{year}{2019}).

\bibitem{glasser2016multi}
\bibinfo{author}{Glasser, M.~F.} \emph{et~al.}
\newblock \bibinfo{journal}{\bibinfo{title}{A multi-modal parcellation of human cerebral cortex}}.
\newblock {\emph{\JournalTitle{Nature}}} \textbf{\bibinfo{volume}{536}}, \bibinfo{pages}{171--178} (\bibinfo{year}{2016}).

\bibitem{yeo2011organization}
\bibinfo{author}{Yeo, B.~T.} \emph{et~al.}
\newblock \bibinfo{journal}{\bibinfo{title}{The organization of the human cerebral cortex estimated by intrinsic functional connectivity}}.
\newblock {\emph{\JournalTitle{Journal of neurophysiology}}}  (\bibinfo{year}{2011}).

\bibitem{biderman2023pythia}
\bibinfo{author}{Biderman, S.} \emph{et~al.}
\newblock \bibinfo{title}{Pythia: A suite for analyzing large language models across training and scaling}.
\newblock In \emph{\bibinfo{booktitle}{International Conference on Machine Learning}}, \bibinfo{pages}{2397--2430} (\bibinfo{organization}{PMLR}, \bibinfo{year}{2023}).

\bibitem{tan2019mixconv}
\bibinfo{author}{Tan, M.} \& \bibinfo{author}{Le, Q.~V.}
\newblock \bibinfo{title}{Mixconv: Mixed depthwise convolutional kernels}.
\newblock In \emph{\bibinfo{booktitle}{BMVC}} (\bibinfo{year}{2019}).

\bibitem{open-llm-leaderboard-v2}
\bibinfo{author}{Fourrier, C.}, \bibinfo{author}{Habib, N.}, \bibinfo{author}{Lozovskaya, A.}, \bibinfo{author}{Szafer, K.} \& \bibinfo{author}{Wolf, T.}
\newblock \bibinfo{title}{Open llm leaderboard v2}.
\newblock \bibinfo{howpublished}{\url{https://huggingface.co/spaces/open-llm-leaderboard/open_llm_leaderboard}} (\bibinfo{year}{2024}).

\bibitem{deng2009imagenet}
\bibinfo{author}{Deng, J.} \emph{et~al.}
\newblock \bibinfo{title}{Imagenet: A large-scale hierarchical image database}.
\newblock In \emph{\bibinfo{booktitle}{2009 IEEE conference on computer vision and pattern recognition}}, \bibinfo{pages}{248--255} (\bibinfo{organization}{Ieee}, \bibinfo{year}{2009}).

\bibitem{lambert2024t}
\bibinfo{author}{Lambert, N.} \emph{et~al.}
\newblock \bibinfo{journal}{\bibinfo{title}{T$\backslash$" ulu 3: Pushing frontiers in open language model post-training}}.
\newblock {\emph{\JournalTitle{arXiv preprint arXiv:2411.15124}}}  (\bibinfo{year}{2024}).

\bibitem{meng2024simpo}
\bibinfo{author}{Meng, Y.}, \bibinfo{author}{Xia, M.} \& \bibinfo{author}{Chen, D.}
\newblock \bibinfo{journal}{\bibinfo{title}{Simpo: Simple preference optimization with a reference-free reward}}.
\newblock {\emph{\JournalTitle{Advances in Neural Information Processing Systems}}} \textbf{\bibinfo{volume}{37}}, \bibinfo{pages}{124198--124235} (\bibinfo{year}{2024}).

\bibitem{zhou2023instruction}
\bibinfo{author}{Zhou, J.} \emph{et~al.}
\newblock \bibinfo{journal}{\bibinfo{title}{Instruction-following evaluation for large language models}}.
\newblock {\emph{\JournalTitle{arXiv preprint arXiv:2311.07911}}}  (\bibinfo{year}{2023}).

\bibitem{suzgun2023challenging}
\bibinfo{author}{Suzgun, M.} \emph{et~al.}
\newblock \bibinfo{title}{Challenging big-bench tasks and whether chain-of-thought can solve them}.
\newblock In \emph{\bibinfo{booktitle}{Findings of the Association for Computational Linguistics: ACL 2023}}, \bibinfo{pages}{13003--13051} (\bibinfo{year}{2023}).

\bibitem{hendrycks2measuring}
\bibinfo{author}{Hendrycks, D.} \emph{et~al.}
\newblock \bibinfo{title}{Measuring mathematical problem solving with the math dataset}.
\newblock In \emph{\bibinfo{booktitle}{Thirty-fifth Conference on Neural Information Processing Systems Datasets and Benchmarks Track (Round 2)}} (\bibinfo{year}{2024}).

\bibitem{rein2024gpqa}
\bibinfo{author}{Rein, D.} \emph{et~al.}
\newblock \bibinfo{title}{Gpqa: A graduate-level google-proof q\&a benchmark}.
\newblock In \emph{\bibinfo{booktitle}{First Conference on Language Modeling}} (\bibinfo{year}{2024}).

\bibitem{sprague2024musr}
\bibinfo{author}{Sprague, Z.}, \bibinfo{author}{Ye, X.}, \bibinfo{author}{Bostrom, K.}, \bibinfo{author}{Chaudhuri, S.} \& \bibinfo{author}{Durrett, G.}
\newblock \bibinfo{journal}{\bibinfo{title}{Musr: Testing the limits of chain-of-thought with multistep soft reasoning}}.
\newblock {\emph{\JournalTitle{ICLR}}}  (\bibinfo{year}{2024}).

\bibitem{wang2024mmlu}
\bibinfo{author}{Wang, Y.} \emph{et~al.}
\newblock \bibinfo{title}{Mmlu-pro: A more robust and challenging multi-task language understanding benchmark}.
\newblock In \emph{\bibinfo{booktitle}{The Thirty-eight Conference on Neural Information Processing Systems Datasets and Benchmarks Track}} (\bibinfo{year}{2024}).

\bibitem{chiang2024chatbot}
\bibinfo{author}{Chiang, W.-L.} \emph{et~al.}
\newblock \bibinfo{title}{Chatbot arena: An open platform for evaluating llms by human preference}.
\newblock In \emph{\bibinfo{booktitle}{Forty-first International Conference on Machine Learning}} (\bibinfo{year}{2024}).

\bibitem{hendrycks2021natural}
\bibinfo{author}{Hendrycks, D.}, \bibinfo{author}{Zhao, K.}, \bibinfo{author}{Basart, S.}, \bibinfo{author}{Steinhardt, J.} \& \bibinfo{author}{Song, D.}
\newblock \bibinfo{title}{Natural adversarial examples}.
\newblock In \emph{\bibinfo{booktitle}{Proceedings of the IEEE/CVF conference on computer vision and pattern recognition}}, \bibinfo{pages}{15262--15271} (\bibinfo{year}{2021}).

\bibitem{hendrycks2021many}
\bibinfo{author}{Hendrycks, D.} \emph{et~al.}
\newblock \bibinfo{title}{The many faces of robustness: A critical analysis of out-of-distribution generalization}.
\newblock In \emph{\bibinfo{booktitle}{Proceedings of the IEEE/CVF international conference on computer vision}}, \bibinfo{pages}{8340--8349} (\bibinfo{year}{2021}).

\bibitem{beyer2020we}
\bibinfo{author}{Beyer, L.}, \bibinfo{author}{H{\'e}naff, O.~J.}, \bibinfo{author}{Kolesnikov, A.}, \bibinfo{author}{Zhai, X.} \& \bibinfo{author}{Oord, A. v.~d.}
\newblock \bibinfo{journal}{\bibinfo{title}{Are we done with imagenet?}}
\newblock {\emph{\JournalTitle{arXiv preprint arXiv:2006.07159}}}  (\bibinfo{year}{2020}).

\bibitem{recht2019imagenet}
\bibinfo{author}{Recht, B.}, \bibinfo{author}{Roelofs, R.}, \bibinfo{author}{Schmidt, L.} \& \bibinfo{author}{Shankar, V.}
\newblock \bibinfo{title}{Do imagenet classifiers generalize to imagenet?}
\newblock In \emph{\bibinfo{booktitle}{International conference on machine learning}}, \bibinfo{pages}{5389--5400} (\bibinfo{organization}{PMLR}, \bibinfo{year}{2019}).

\bibitem{wang2019learning}
\bibinfo{author}{Wang, H.}, \bibinfo{author}{Ge, S.}, \bibinfo{author}{Lipton, Z.} \& \bibinfo{author}{Xing, E.~P.}
\newblock \bibinfo{journal}{\bibinfo{title}{Learning robust global representations by penalizing local predictive power}}.
\newblock {\emph{\JournalTitle{Advances in neural information processing systems}}} \textbf{\bibinfo{volume}{32}} (\bibinfo{year}{2019}).

\bibitem{mcinnes2018umap}
\bibinfo{author}{McInnes, L.}, \bibinfo{author}{Healy, J.}, \bibinfo{author}{Saul, N.} \& \bibinfo{author}{Gro{\ss}berger, L.}
\newblock \bibinfo{journal}{\bibinfo{title}{Umap: Uniform manifold approximation and projection}}.
\newblock {\emph{\JournalTitle{Journal of Open Source Software}}} \textbf{\bibinfo{volume}{3}}, \bibinfo{pages}{861} (\bibinfo{year}{2018}).

\bibitem{paperno2016lambada}
\bibinfo{author}{Paperno, D.} \emph{et~al.}
\newblock \bibinfo{title}{The lambada dataset: Word prediction requiring a broad discourse context}.
\newblock In \emph{\bibinfo{booktitle}{Proceedings of the 54th Annual Meeting of the Association for Computational Linguistics (Volume 1: Long Papers)}}, \bibinfo{pages}{1525--1534} (\bibinfo{year}{2016}).

\bibitem{bisk2020piqa}
\bibinfo{author}{Bisk, Y.}, \bibinfo{author}{Zellers, R.}, \bibinfo{author}{Gao, J.}, \bibinfo{author}{Choi, Y.} \emph{et~al.}
\newblock \bibinfo{title}{Piqa: Reasoning about physical commonsense in natural language}.
\newblock In \emph{\bibinfo{booktitle}{Proceedings of the AAAI conference on artificial intelligence}}, vol.~\bibinfo{volume}{34}, \bibinfo{pages}{7432--7439} (\bibinfo{year}{2020}).

\bibitem{simonyan2014very}
\bibinfo{author}{Simonyan, K.} \& \bibinfo{author}{Zisserman, A.}
\newblock \bibinfo{journal}{\bibinfo{title}{Very deep convolutional networks for large-scale image recognition}}.
\newblock {\emph{\JournalTitle{arXiv preprint arXiv:1409.1556}}}  (\bibinfo{year}{2014}).

\bibitem{chollet2017xception}
\bibinfo{author}{Chollet, F.}
\newblock \bibinfo{title}{Xception: Deep learning with depthwise separable convolutions}.
\newblock In \emph{\bibinfo{booktitle}{Proceedings of the IEEE conference on computer vision and pattern recognition}}, \bibinfo{pages}{1251--1258} (\bibinfo{year}{2017}).

\bibitem{he2016deep}
\bibinfo{author}{He, K.}, \bibinfo{author}{Zhang, X.}, \bibinfo{author}{Ren, S.} \& \bibinfo{author}{Sun, J.}
\newblock \bibinfo{title}{Deep residual learning for image recognition}.
\newblock In \emph{\bibinfo{booktitle}{Proceedings of the IEEE conference on computer vision and pattern recognition}}, \bibinfo{pages}{770--778} (\bibinfo{year}{2016}).

\bibitem{koonce2021mobilenetv3}
\bibinfo{author}{Koonce, B.} \& \bibinfo{author}{Koonce, B.}
\newblock \bibinfo{journal}{\bibinfo{title}{Mobilenetv3}}.
\newblock {\emph{\JournalTitle{Convolutional Neural Networks with Swift for Tensorflow: Image Recognition and Dataset Categorization}}} \bibinfo{pages}{125--144} (\bibinfo{year}{2021}).

\bibitem{koonce2021efficientnet}
\bibinfo{author}{Koonce, B.}
\newblock \bibinfo{title}{Efficientnet}.
\newblock In \emph{\bibinfo{booktitle}{Convolutional neural networks with swift for Tensorflow: image recognition and dataset categorization}}, \bibinfo{pages}{109--123} (\bibinfo{publisher}{Springer}, \bibinfo{year}{2021}).

\bibitem{liu2022convnet}
\bibinfo{author}{Liu, Z.} \emph{et~al.}
\newblock \bibinfo{title}{A convnet for the 2020s}.
\newblock In \emph{\bibinfo{booktitle}{Proceedings of the IEEE/CVF conference on computer vision and pattern recognition}}, \bibinfo{pages}{11976--11986} (\bibinfo{year}{2022}).

\bibitem{dai2021coatnet}
\bibinfo{author}{Dai, Z.}, \bibinfo{author}{Liu, H.}, \bibinfo{author}{Le, Q.~V.} \& \bibinfo{author}{Tan, M.}
\newblock \bibinfo{journal}{\bibinfo{title}{Coatnet: Marrying convolution and attention for all data sizes}}.
\newblock {\emph{\JournalTitle{Advances in neural information processing systems}}} \textbf{\bibinfo{volume}{34}}, \bibinfo{pages}{3965--3977} (\bibinfo{year}{2021}).

\bibitem{yang2022focal}
\bibinfo{author}{Yang, J.}, \bibinfo{author}{Li, C.}, \bibinfo{author}{Dai, X.} \& \bibinfo{author}{Gao, J.}
\newblock \bibinfo{journal}{\bibinfo{title}{Focal modulation networks}}.
\newblock {\emph{\JournalTitle{Advances in Neural Information Processing Systems}}} \textbf{\bibinfo{volume}{35}}, \bibinfo{pages}{4203--4217} (\bibinfo{year}{2022}).

\bibitem{hasson2020direct}
\bibinfo{author}{Hasson, U.}, \bibinfo{author}{Nastase, S.~A.} \& \bibinfo{author}{Goldstein, A.}
\newblock \bibinfo{journal}{\bibinfo{title}{Direct fit to nature: an evolutionary perspective on biological and artificial neural networks}}.
\newblock {\emph{\JournalTitle{Neuron}}} \textbf{\bibinfo{volume}{105}}, \bibinfo{pages}{416--434} (\bibinfo{year}{2020}).

\bibitem{marblestone2016toward}
\bibinfo{author}{Marblestone, A.~H.}, \bibinfo{author}{Wayne, G.} \& \bibinfo{author}{Kording, K.~P.}
\newblock \bibinfo{journal}{\bibinfo{title}{Toward an integration of deep learning and neuroscience}}.
\newblock {\emph{\JournalTitle{Frontiers in computational neuroscience}}} \textbf{\bibinfo{volume}{10}}, \bibinfo{pages}{94} (\bibinfo{year}{2016}).

\bibitem{caucheteux2022brains}
\bibinfo{author}{Caucheteux, C.} \& \bibinfo{author}{King, J.-R.}
\newblock \bibinfo{journal}{\bibinfo{title}{Brains and algorithms partially converge in natural language processing}}.
\newblock {\emph{\JournalTitle{Communications biology}}} \textbf{\bibinfo{volume}{5}}, \bibinfo{pages}{134} (\bibinfo{year}{2022}).

\bibitem{fedorenko2014reworking}
\bibinfo{author}{Fedorenko, E.} \& \bibinfo{author}{Thompson-Schill, S.~L.}
\newblock \bibinfo{journal}{\bibinfo{title}{Reworking the language network}}.
\newblock {\emph{\JournalTitle{Trends in cognitive sciences}}} \textbf{\bibinfo{volume}{18}}, \bibinfo{pages}{120--126} (\bibinfo{year}{2014}).

\bibitem{pereira2018toward}
\bibinfo{author}{Pereira, F.} \emph{et~al.}
\newblock \bibinfo{journal}{\bibinfo{title}{Toward a universal decoder of linguistic meaning from brain activation}}.
\newblock {\emph{\JournalTitle{Nature communications}}} \textbf{\bibinfo{volume}{9}}, \bibinfo{pages}{963} (\bibinfo{year}{2018}).

\bibitem{cichy2016comparison}
\bibinfo{author}{Cichy, R.~M.}, \bibinfo{author}{Khosla, A.}, \bibinfo{author}{Pantazis, D.}, \bibinfo{author}{Torralba, A.} \& \bibinfo{author}{Oliva, A.}
\newblock \bibinfo{journal}{\bibinfo{title}{Comparison of deep neural networks to spatio-temporal cortical dynamics of human visual object recognition reveals hierarchical correspondence}}.
\newblock {\emph{\JournalTitle{Scientific reports}}} \textbf{\bibinfo{volume}{6}}, \bibinfo{pages}{27755} (\bibinfo{year}{2016}).

\bibitem{goldstein2022shared}
\bibinfo{author}{Goldstein, A.} \emph{et~al.}
\newblock \bibinfo{journal}{\bibinfo{title}{Shared computational principles for language processing in humans and deep language models}}.
\newblock {\emph{\JournalTitle{Nature neuroscience}}} \textbf{\bibinfo{volume}{25}}, \bibinfo{pages}{369--380} (\bibinfo{year}{2022}).

\bibitem{dobs2022brain}
\bibinfo{author}{Dobs, K.}, \bibinfo{author}{Martinez, J.}, \bibinfo{author}{Kell, A.~J.} \& \bibinfo{author}{Kanwisher, N.}
\newblock \bibinfo{journal}{\bibinfo{title}{Brain-like functional specialization emerges spontaneously in deep neural networks}}.
\newblock {\emph{\JournalTitle{Science advances}}} \textbf{\bibinfo{volume}{8}}, \bibinfo{pages}{eabl8913} (\bibinfo{year}{2022}).

\bibitem{binder2016toward}
\bibinfo{author}{Binder, J.~R.} \emph{et~al.}
\newblock \bibinfo{journal}{\bibinfo{title}{Toward a brain-based componential semantic representation}}.
\newblock {\emph{\JournalTitle{Cognitive neuropsychology}}} \textbf{\bibinfo{volume}{33}}, \bibinfo{pages}{130--174} (\bibinfo{year}{2016}).

\bibitem{huth2016natural}
\bibinfo{author}{Huth, A.~G.}, \bibinfo{author}{De~Heer, W.~A.}, \bibinfo{author}{Griffiths, T.~L.}, \bibinfo{author}{Theunissen, F.~E.} \& \bibinfo{author}{Gallant, J.~L.}
\newblock \bibinfo{journal}{\bibinfo{title}{Natural speech reveals the semantic maps that tile human cerebral cortex}}.
\newblock {\emph{\JournalTitle{Nature}}} \textbf{\bibinfo{volume}{532}}, \bibinfo{pages}{453--458} (\bibinfo{year}{2016}).

\bibitem{margulies2016situating}
\bibinfo{author}{Margulies, D.~S.} \emph{et~al.}
\newblock \bibinfo{journal}{\bibinfo{title}{Situating the default-mode network along a principal gradient of macroscale cortical organization}}.
\newblock {\emph{\JournalTitle{Proceedings of the National Academy of Sciences}}} \textbf{\bibinfo{volume}{113}}, \bibinfo{pages}{12574--12579} (\bibinfo{year}{2016}).

\bibitem{huntenburg2018large}
\bibinfo{author}{Huntenburg, J.~M.}, \bibinfo{author}{Bazin, P.-L.} \& \bibinfo{author}{Margulies, D.~S.}
\newblock \bibinfo{journal}{\bibinfo{title}{Large-scale gradients in human cortical organization}}.
\newblock {\emph{\JournalTitle{Trends in cognitive sciences}}} \textbf{\bibinfo{volume}{22}}, \bibinfo{pages}{21--31} (\bibinfo{year}{2018}).

\bibitem{fedorenko2016language}
\bibinfo{author}{Fedorenko, E.} \& \bibinfo{author}{Varley, R.}
\newblock \bibinfo{journal}{\bibinfo{title}{Language and thought are not the same thing: evidence from neuroimaging and neurological patients}}.
\newblock {\emph{\JournalTitle{Annals of the New York Academy of Sciences}}} \textbf{\bibinfo{volume}{1369}}, \bibinfo{pages}{132--153} (\bibinfo{year}{2016}).

\bibitem{hagoort2019neurobiology}
\bibinfo{author}{Hagoort, P.}
\newblock \bibinfo{journal}{\bibinfo{title}{The neurobiology of language beyond single-word processing}}.
\newblock {\emph{\JournalTitle{Science}}} \textbf{\bibinfo{volume}{366}}, \bibinfo{pages}{55--58} (\bibinfo{year}{2019}).

\bibitem{saxe2019mathematical}
\bibinfo{author}{Saxe, A.~M.}, \bibinfo{author}{McClelland, J.~L.} \& \bibinfo{author}{Ganguli, S.}
\newblock \bibinfo{journal}{\bibinfo{title}{A mathematical theory of semantic development in deep neural networks}}.
\newblock {\emph{\JournalTitle{Proceedings of the National Academy of Sciences}}} \textbf{\bibinfo{volume}{116}}, \bibinfo{pages}{11537--11546} (\bibinfo{year}{2019}).

\bibitem{kriegeskorte2015deep}
\bibinfo{author}{Kriegeskorte, N.}
\newblock \bibinfo{journal}{\bibinfo{title}{Deep neural networks: a new framework for modeling biological vision and brain information processing}}.
\newblock {\emph{\JournalTitle{Annual review of vision science}}} \textbf{\bibinfo{volume}{1}}, \bibinfo{pages}{417--446} (\bibinfo{year}{2015}).

\bibitem{kriegeskorte2018cognitive}
\bibinfo{author}{Kriegeskorte, N.} \& \bibinfo{author}{Douglas, P.~K.}
\newblock \bibinfo{journal}{\bibinfo{title}{Cognitive computational neuroscience}}.
\newblock {\emph{\JournalTitle{Nature neuroscience}}} \textbf{\bibinfo{volume}{21}}, \bibinfo{pages}{1148--1160} (\bibinfo{year}{2018}).

\bibitem{kietzmann2019recurrence}
\bibinfo{author}{Kietzmann, T.~C.} \emph{et~al.}
\newblock \bibinfo{journal}{\bibinfo{title}{Recurrence is required to capture the representational dynamics of the human visual system}}.
\newblock {\emph{\JournalTitle{Proceedings of the National Academy of Sciences}}} \textbf{\bibinfo{volume}{116}}, \bibinfo{pages}{21854--21863} (\bibinfo{year}{2019}).

\bibitem{kubilius2019brain}
\bibinfo{author}{Kubilius, J.} \emph{et~al.}
\newblock \bibinfo{journal}{\bibinfo{title}{Brain-like object recognition with high-performing shallow recurrent anns}}.
\newblock {\emph{\JournalTitle{Advances in neural information processing systems}}} \textbf{\bibinfo{volume}{32}} (\bibinfo{year}{2019}).

\bibitem{dodge2021documenting}
\bibinfo{author}{Dodge, J.} \emph{et~al.}
\newblock \bibinfo{journal}{\bibinfo{title}{Documenting large webtext corpora: A case study on the colossal clean crawled corpus}}.
\newblock {\emph{\JournalTitle{arXiv preprint arXiv:2104.08758}}}  (\bibinfo{year}{2021}).

\bibitem{bender2021dangers}
\bibinfo{author}{Bender, E.~M.}, \bibinfo{author}{Gebru, T.}, \bibinfo{author}{McMillan-Major, A.} \& \bibinfo{author}{Shmitchell, S.}
\newblock \bibinfo{title}{On the dangers of stochastic parrots: Can language models be too big?}
\newblock In \emph{\bibinfo{booktitle}{Proceedings of the 2021 ACM conference on fairness, accountability, and transparency}}, \bibinfo{pages}{610--623} (\bibinfo{year}{2021}).

\bibitem{kriegeskorte2019interpreting}
\bibinfo{author}{Kriegeskorte, N.} \& \bibinfo{author}{Douglas, P.~K.}
\newblock \bibinfo{journal}{\bibinfo{title}{Interpreting encoding and decoding models}}.
\newblock {\emph{\JournalTitle{Current opinion in neurobiology}}} \textbf{\bibinfo{volume}{55}}, \bibinfo{pages}{167--179} (\bibinfo{year}{2019}).

\bibitem{lindsay2021convolutional}
\bibinfo{author}{Lindsay, G.~W.}
\newblock \bibinfo{journal}{\bibinfo{title}{Convolutional neural networks as a model of the visual system: Past, present, and future}}.
\newblock {\emph{\JournalTitle{Journal of cognitive neuroscience}}} \textbf{\bibinfo{volume}{33}}, \bibinfo{pages}{2017--2031} (\bibinfo{year}{2021}).

\bibitem{jiangMistral7B2023}
\bibinfo{author}{Jiang, A.~Q.} \emph{et~al.}
\newblock \bibinfo{title}{Mistral {{7B}}}, \doiprefix\url{10.48550/arXiv.2310.06825}.
\newblock \eprint{2310.06825}.

\bibitem{abdin2024phi}
\bibinfo{author}{Abdin, M.} \emph{et~al.}
\newblock \bibinfo{journal}{\bibinfo{title}{Phi-3 technical report: A highly capable language model locally on your phone}}.
\newblock {\emph{\JournalTitle{arXiv preprint arXiv:2404.14219}}}  (\bibinfo{year}{2024}).

\bibitem{team2023internlm}
\bibinfo{author}{Team, I.}
\newblock \bibinfo{title}{Internlm: A multilingual language model with progressively enhanced capabilities} (\bibinfo{year}{2023}).

\bibitem{wolf-etal-2020-transformers}
\bibinfo{author}{Wolf, T.} \emph{et~al.}
\newblock \bibinfo{title}{Transformers: State-of-the-art natural language processing}.
\newblock In \emph{\bibinfo{booktitle}{Proceedings of the 2020 Conference on Empirical Methods in Natural Language Processing: System Demonstrations}}, \bibinfo{pages}{38--45} (\bibinfo{publisher}{Association for Computational Linguistics}, \bibinfo{address}{Online}, \bibinfo{year}{2020}).

\bibitem{devlin2019bert}
\bibinfo{author}{Devlin, J.}, \bibinfo{author}{Chang, M.-W.}, \bibinfo{author}{Lee, K.} \& \bibinfo{author}{Toutanova, K.}
\newblock \bibinfo{title}{Bert: Pre-training of deep bidirectional transformers for language understanding}.
\newblock In \emph{\bibinfo{booktitle}{Proceedings of the 2019 conference of the North American chapter of the association for computational linguistics: human language technologies, volume 1 (long and short papers)}}, \bibinfo{pages}{4171--4186} (\bibinfo{year}{2019}).

\bibitem{liu2019roberta}
\bibinfo{author}{Liu, Y.} \emph{et~al.}
\newblock \bibinfo{journal}{\bibinfo{title}{Roberta: A robustly optimized bert pretraining approach}}.
\newblock {\emph{\JournalTitle{arXiv preprint arXiv:1907.11692}}}  (\bibinfo{year}{2019}).

\bibitem{rw2019timm}
\bibinfo{author}{Wightman, R.}
\newblock \bibinfo{title}{Pytorch image models}.
\newblock \bibinfo{howpublished}{\url{https://github.com/rwightman/pytorch-image-models}}, \doiprefix\url{10.5281/zenodo.4414861} (\bibinfo{year}{2019}).

\bibitem{byrge2019high}
\bibinfo{author}{Byrge, L.} \& \bibinfo{author}{Kennedy, D.~P.}
\newblock \bibinfo{journal}{\bibinfo{title}{High-accuracy individual identification using a “thin slice” of the functional connectome}}.
\newblock {\emph{\JournalTitle{Network Neuroscience}}} \textbf{\bibinfo{volume}{3}}, \bibinfo{pages}{363--383} (\bibinfo{year}{2019}).

\bibitem{ward1963hierarchical}
\bibinfo{author}{Ward~Jr, J.~H.}
\newblock \bibinfo{journal}{\bibinfo{title}{Hierarchical grouping to optimize an objective function}}.
\newblock {\emph{\JournalTitle{Journal of the American statistical association}}} \textbf{\bibinfo{volume}{58}}, \bibinfo{pages}{236--244} (\bibinfo{year}{1963}).

\end{thebibliography}







\section*{Author contributions statement}

G.S. performed the primary analyses, and implemented the computational framework. D.Z. contributed to the conceptualization of the research idea. G.S., D.Z., and Q.Z. jointly designed the experiments, established the analytical pipeline, and developed the methodological approach. Y.D. assisted with data processing and analysis. Y.Z. supervised the project and provided funding support. G.S. wrote the original draft. All authors contributed to reviewing and editing the manuscript. Y.Z. provided critical revisions of the manuscript and approved the final version.


\section*{Data Availability}


All brain imaging data used in this study is from the Natural Scenes Dataset (NSD) (https://naturalscenesdataset.org/)~\cite{allen2022massive}, a large-scale fMRI dataset containing neural responses to natural images. The image stimuli and captions were sourced from the COCO dataset (https://cocodataset.org/)~\cite{lin2014microsoft}. For language model analyses, we used publicly available pre-trained models from the HuggingFace Transformers library (https://huggingface.co/models)~\cite{wolf-etal-2020-transformers}, including models from the Qwen~\cite{yang2024qwen2}, Llama~\cite{grattafiori2024llama}, Gemma~\cite{team2024gemma}, Mistral~\cite{jiangMistral7B2023}, Phi~\cite{abdin2024phi} families and others~\cite{team2023internlm, lambert2024t, meng2024simpo, biderman2023pythia}. For vision model analyses, we used pre-trained models from the TIMM library (https://github.com/huggingface/pytorch-image-models)~\cite{rw2019timm}. Model performance metrics were obtained from LLM Leaderboard 2 (https://huggingface.co/spaces/HuggingFaceH4/open\_llm\_leaderboard)~\cite{open-llm-leaderboard-v2} for language models and standard ImageNet~\cite{krizhevsky2012imagenet} benchmarks for vision models. For longitudinal analyses, we used checkpoints from the Pythia model family (https://github.com/EleutherAI/pythia)~\cite{biderman2023pythia} and trained MixNet models~\cite{tan2019mixconv} with saved checkpoints.

\section*{Code Availability}

All code used for data processing, model representation extraction, alignment computation, and statistical analyses is available on \href{GitHub}{https://github.com/FloyedShen/BrainAlign}. 







\end{document}